\documentclass[11pt]{article}
\usepackage{fullpage}
\usepackage{latexsym}

\title{The Random Oracle Methodology, Revisited\thanks{Extended
abstract has appeared in the {\em Proc. of the 30th ACM Symp. on
Theory of Computing (STOC)}, pages 209--218, 1998.}}

\def\IBM{IBM Watson, P.O. Box 704, Yorktown Height, NY 10598, USA}

\def\ROM{Random Oracle Model}

\def\corInt{correlation intractable}
\def\CorInty{Correlation intractability}
\def\corInty{correlation intractability}

\author{Ran Canetti\thanks{\IBM. E-mail: {\tt canetti@watson.ibm.com}}  
\and Oded Goldreich\thanks{Department of Computer Science,  
Weizmann Institute of Science, Rehovot, {\sc Israel}. 
E-mail: {\tt oded@wisdom.weizmann.ac.il}. 
Work done while visiting LCS, MIT.
Partially supported by DARPA grant DABT63-96-C-0018.} 
\and Shai Halevi\thanks{\IBM. E-mail: {\tt shaih@watson.ibm.com}}} 

\def\ni{\noindent}
\def\etal{{\it et.~al.}}

\newcommand{\eqdef}{\stackrel{\rm def}{=}} 
\newcommand{\bitset}{\{0,1\}}

\newcommand{\negl}{{\rm negl}}
\newcommand{\sk}{{\rm sk}}
\newcommand{\vk}{{\rm vk}}
\newcommand{\ek}{{\rm ek}}
\newcommand{\dk}{{\rm dk}}
\newcommand{\msg}{{\rm msg}}
\def\th{{\rm th}}
\newcommand{\ov}{\overline}
\newcommand{\e}{\epsilon}
\newcommand{\f}{{\cal O}}

\newtheorem{thm}{Theorem}[section]      % A counter for Theorems etc
\newcommand{\BT}{\begin{thm}}   
\newcommand{\ET}{\end{thm}}

\newtheorem{dfn}[thm]{Definition}      % 
\newcommand{\BD}{\begin{dfn}}   
\newcommand{\ED}{\end{dfn}}

\newtheorem{propos}[thm]{Proposition}      % 
\newcommand{\BP}{\begin{propos}}   
\newcommand{\EP}{\end{propos}}

\newtheorem{corr}[thm]{Corollary}      % 
\newcommand{\BCR}{\begin{corr}} 
\newcommand{\ECR}{\end{corr}}

\newtheorem{rmk}[thm]{Remark}      % 
\newcommand{\BRK}{\begin{rmk}\rm\ } 
\newcommand{\ERK}{\end{rmk}}
\newcommand{\BR}[1]{\begin{rmk}[#1] \rm\ } 
\newcommand{\ER}{\end{rmk}}
\newtheorem{Ithm}{Informal Theorem}[section]  % A counter for Theorems in Intro
\newcommand{\BIT}{\begin{Ithm}}   
\newcommand{\EIT}{\end{Ithm}}

\newtheorem{Idfn}[Ithm]{Informal Definition}      % 
\newcommand{\BID}{\begin{Idfn}}   
\newcommand{\EID}{\end{Idfn}}
\newcommand{\BE}{\begin{enumerate}}
\newcommand{\EE}{\end{enumerate}}

\newcommand{\BI}{\begin{itemize}}
\newcommand{\EI}{\end{itemize}}

\def\FullBox{\hbox{\vrule width 8pt height 8pt depth 0pt}}
\newcommand{\qed}{\;\;\;\FullBox}

\newenvironment{proof}{\noindent{\bf Proof:~~}}{\(\qed\)}
\newcommand{\BPF}{\begin{proof}} 
\newcommand {\EPF}{\end{proof}}

\newcommand{\qedsketch}{\;\;\;\Box}

\def\qqed{$\Box$}
\newcommand{\dlen}{\ell_{\rm in}} 
\newcommand{\rlen}{\ell_{\rm out}}

\newcommand{\N}{{\sf N}} 
\newcommand{\intfnc}{{\N\!\to\!\N}}

\newcommand{\prob}{{\rm Pr}} 
\newcommand{\poly}{{\rm poly}}

\newcommand{\vect}[1]{\langle{#1}\rangle}

\newcommand{\eqref}[1]{Eq.~(\ref{#1})}
\newcommand{\secref}[1]{Section~\ref{#1}}

\newcommand{\defref}[1]{Definition~\ref{#1}}
\newcommand{\thmref}[1]{Theorem~\ref{#1}}

\newcommand{\prpref}[1]{Proposition~\ref{#1}}

\begin{document} 

\begin{titlepage}
\maketitle

\begin{abstract} 
We take a critical look at the relationship between the 
security of cryptographic schemes in the Random Oracle Model, and 
the security of the schemes that result from implementing the random 
oracle by so called ``cryptographic hash functions''. 

The main result of this paper is a negative one: There exist signature 
and encryption schemes that are secure in the Random Oracle Model, but 
for which {\em any implementation}\/ of the random oracle results in 
insecure schemes. 

In the process of devising the above schemes, we consider possible 
definitions for the notion of a ``good implementation'' of a random 
oracle, pointing out limitations and challenges.
\end{abstract}

\vfill
\BI
\item Cryptography (Encryption and Signature Schemes, The Random Oracle model); 
\item Complexity Theory (diagonalization, application of CS-Proofs).  
\EI 
\medskip\ni{~}
\end{titlepage}

%%%%%%%%%%%%%%%%%%%%%%%%%%%%%%%%%%%%%%%%%%%%%%%%%%%%%%%%%%%%%%
\section{Introduction}\label{intro.sec}
%%%%%%%%%%%%%%%%%%%%%%%%%%%%%%%%%%%%%%%%%%%%%%%%%%%%%%%%%%%%%%
A popular methodology for designing cryptographic protocols consists 
of the following two steps. One first designs an {\em ideal}\/ system 
in which all parties (including the adversary) have oracle access to 
a truly random function, and proves the security of this ideal system. 
Next, one replaces the random oracle by a ``good cryptographic hashing 
function'' (such as MD5 or SHA), providing all parties (including the 
adversary) with a succinct description of this function. Thus, one 
obtains an {\em implementation}\/ of the ideal system in a ``real-world'' 
where random oracles do not exist. 
This methodology, explicitly formulated by Bellare and Rogaway~\cite{BeRo93} 
and hereafter referred to as the {\em random oracle methodology}, 
has been used in many works (see, for example, 
\cite{FiSh86,Sc91,GuQu88,Ok92,BeRo93,Mi94,BeRo96,PoSt96}). 

Although the random oracle methodology seems to be useful in practice, 
it is unclear how to put this methodology on firm grounds.  One can 
indeed make clear statements regarding the operation of the ideal system, 
but it is not clear what happens when one replaces the random oracle by 
a function that has a succinct description available to all parties. 
What one would have liked is (at least a definition of) a class of functions 
that, when used to replace the random oracle, maintains the security of the 
ideal scheme. The purpose of this work is to point out fundamental 
difficulties in proceeding towards this goal. We demonstrate that 
the traditional approach of providing a {\em single} robust definition 
that supports a wide range of applications is bound to fail. 
That is, one  cannot expect to see definitions such as of pseudorandom 
generators or functions~\cite{BlMi84,yao82a,ggm86}, and general 
results of the type saying that these can be used in any application 
in which parties are restricted merely by computing resources. 
Specifically, we identify a specific property
of the random oracle, that
seems to capture one aspect of the random oracle methodology 
(and in particular seems to underline heuristics such as the 
Fiat--Shamir transformation of a three-round identification scheme 
into a signature scheme in the ~\cite{FiSh86}). 
We show that even a minimalistic formulation of this property, 
called {\sf correlation intractability}, 
cannot be obtained by any fully specified 
function (or function ensemble). 

To demonstrate the implications of the above to the security of 
cryptographic systems, we show that systems whose security relies 
on the ``correlation intractability'' of their oracle may be secure 
in the \ROM, and yet be insecure when implemented using any fully 
specified function (or function ensemble). In particular, we describe 
schemes for digital signatures and public-key encryption that are 
secure in the \ROM, but for which any implementation yields insecure 
schemes. This refutes the belief that a security proof in the \ROM\ 
means that there are no ``structural flaws'' in the scheme.

\subsection{The Setting}
%--------------------------
For the purpose of the following discussion, a cryptographic system consists
of a set of parties, which are 
modeled by probabilistic polynomial time interactive Turing machines.
A {\sf cryptographic application} 
comes with a security requirement specifying 
the adversary's abilities and when the latter is considered successful. 
The abilities of the adversary include its computational power 
(typically, an arbitrary polynomial-time machine) and the ways in which 
it can interact with the other parties. 
The {\em success} of the adversary is defined by means of a predetermined 
polynomial-time predicate of the {\em application's global view}.%
\footnote{\ 
The application's global view consists of the initial inputs of all 
the parties (including the adversary), their internal coin tosses, and 
all the messages which were exchanged among them.
}\ 
A system is considered {\sf secure} if any adversary with the 
given abilities has only a negligible probability of success. 

\subsubsection{The Random Oracle Model} 
In a scheme that operates in the \ROM, all parties (including the 
adversary) interact with one another as usual 
interactive machines, but in addition they can make oracle queries. It 
is postulated that all oracle queries, regardless of the identity of 
the party making them, are answered by a single function, denoted $\f$, 
that is uniformly selected among all possible functions. The set of 
possible functions is determined by a length function, $\rlen(\cdot)$,
and by the security parameter of the system. Specifically, given 
security parameter $k$ we consider functions mapping 
$\bitset^{{\rm poly}(k)}$ to $\bitset^{\rlen(k)}$.
A set of interactive oracle machines as above corresponds to an 
{\em ideal system for one specific application}. 
Security of an ideal system is defined as usual. That is,
an ideal system is considered secure if any adversary with the 
given abilities (including oracle access)
has only a negligible probability of success. Here the probability is
taken also over the choices of the random oracle.

\subsubsection{Implementing an ideal system}
Loosely speaking, by ``implementing'' a particular ideal system we mean 
using an easy-to-evaluate function $f$ instead of the random oracle. 
That is, whenever the ideal system queries the oracle with a value 
$x$, the implementation instead evaluates $f(x)$. 
Formally defining this notion, however, takes some care. Below we 
briefly examine (and discard of) the notion of implementation by a 
single function, and then present the notion of implementation by a 
function ensemble, which is the notion we use throughout the paper.

\paragraph{Implementation by a single function.} 
In accordance with the above discussion, 
each ideal system (for some specific application), $\Pi$, is 
transformed into a real system (for the same application) by
transforming each interactive oracle machine, 
into a standard interactive machine in the natural manner.
That is, each oracle call is replaced by the evaluation of
a fixed function $f$ on the corresponding query.% 
\footnote{\ Formally, the function $f$ also takes as input the 
security parameter $k$, so that the function $f_k(\cdot) \eqdef 
f(k,\cdot)$ maps $\bitset^{{\rm poly}(k)}$ to $\bitset^{\rlen(k)}$.}

The above system is called an {\em implementation of $\Pi$ using function 
$f$}. The adversary, attacking this implementation, may mimic the behavior 
of the adversary of the ideal system, by evaluating $f$ at arguments of 
its choice,  but it needs not do so. 
In particular, it may obtain some global insight into the structure of 
the function $f$, and use this insight towards its vicious goals. An 
implementation is called {\sf secure} if any adversary attacking it may 
succeed only with negligible probability, where the success event is defined 
exactly as in the ideal system (i.e., it is defined by the same 
polynomial-time computable predicate of the application's global view). 

Using this notion of an implementation, we would like to say that a 
function $f$ is a ``good implementation of a random oracle'' if for any 
ideal system $\Pi$, security of $\Pi$ implies security of the
implementation of $\Pi$ using $f$. 
It is very easy to see, however, that no (single) polynomial-time 
computable function can provide a good implementation of a random oracle. 
Consider, for example, a candidate function $f$. Then, a (contrived) 
application for which $f$ does not provide a good implementation 
consists of an oracle machine (representing an honest 
party) that upon receiving a message $m$, makes query $m$ to the oracle 
and reveals its private input if the oracle answers with $f(m)$. 
Suppose that the adversary is deemed successful whenever the honest party 
reveals its private input. Clearly, this ideal system is secure (in the 
Random Oracle Model), since the random oracle will return the value
$f(m)$ only with negligible probability; 
however, its implementation using $f$ is certainly not secure.

\paragraph{Implementation by a function ensemble.} 
In face of the failure of the above naive attempt, 
a more sophisticated interpretation is indeed called for. 
Here one considers the substitution of the random oracle by a function 
randomly selected from a collection of functions. In this setting, we 
have a ``system set-up'' phase, in which the function is selected once 
and for all, and its description is available to all parties.%
\footnote{\ 
In the sequel we consider examples of public key signature and 
encryption schemes. In these schemes, the initialization (set-up)
step is combined with the key-generation step of the original scheme.
}\ 
After this set-up phase, this function is used in place of the 
random oracle just as above. 
A little more precisely, 
we consider a function ensemble ${\cal F} = \{F_k | k\in\N\}$, where 
$$
F_k = \{
f_s\!:\!\bitset^{{\rm poly}(k)}\!\to\!\bitset^{\rlen(k)}
\}_{s\in\bitset^k}\;,
$$ 
such that there exists a polynomial time algorithm 
that, on input $s$ and $x$, returns $f_s(x)$. 
The implementation of an ideal system, $\Pi$, by the function ensemble 
$\cal F$ is obtained as follows. On security parameter $k$, we uniformly 
select $s\in\bitset^k$, and make $s$ available to all parties including 
the adversary. 
Given this initialization phase, we replace each oracle call of 
an interactive oracle machine by the evaluation of the function $f_s$ 
on the corresponding query. The resulting system is called an {\em 
implementation of $\Pi$ using function ensemble $\cal F$}.

Again, the adversary may (but need not necessarily) mimic the behavior of 
the adversary in the Random Oracle Model by evaluating $f_s$ at arguments 
of its choice. 
Such a real system is called {\sf secure} if any adversary attacking it 
has only a negligible probability of success, where the probability is taken 
over the random choice of $s$ as well as the coins of all the parties. 
As before, we would like to say that an ensemble ${\cal F}$ provides a 
``good implementation of a random oracle'' if for every ideal system $\Pi$, 
if $\Pi$ is secure then so is the implementation of $\Pi$ using $\cal F$. 
Notice that in this case, the contrived example from above does not work 
anymore, since the success event must be independent of the random choice 
of $s$. Nonetheless, this work implies that 
no function ensemble can provide a good implementation of a random oracle. 
We elaborate in the next subsection.

\subsection{Our Results}
%-----------------------
\subsubsection{Correlation intractability.}
One property we certainly expect from a good implementation of a random 
oracle is that it should be infeasible to find inputs to the function 
that stand in some ``rare'' relationship with the corresponding outputs.
Indeed, many applications of the random-oracle methodology (such as 
the Fiat-Shamir heuristic) assume that it is infeasible to find an 
input-output pair that stands in a particular relations induced by 
the application. 
Trying to formulate this property, we may require that given the description 
of the function it is hard to find a sequence of preimages that together 
with their images (under this function) satisfy some given relation. 
Clearly, this can only hold for relations for which finding such sequences 
is hard in the \ROM. %\ (i.e., w.r.t a random oracle). 
That is, {\sc if} it is hard to find a sequence of 
preimages that together with their images under a random oracle satisfy 
relation $R$, {\sc then} given the description of a ``good'' function $f_s$ 
it should be hard to find a sequence of preimages that together with their 
images under $f_s$ satisfy $R$. 

In fact, we mainly consider the task of 
finding a {\em single} 
preimage that together with its image satisfies some property. Loosely 
speaking, a relation is called {\sf evasive} if when given access to a 
random oracle $\f$, it is infeasible to find a string $x$ so that the pair 
$(x,\f(x))$ is in the relation. (For instance, the relation
$\{(x,0^{\rlen(k)}):x\in\bitset^*\}$ is evasive. The relation
$\{(x,0y):x\in\bitset^*,y\in\bitset^{\rlen(k)-1}\}$ is not.) 
A function ensemble $\cal F$ (as above) is called {\sf correlation 
intractable} if for every evasive relation, given the description of 
a uniformly selected function $f_s\in F_k$ it is infeasible to find an 
$x$ such that $(x,f_s(x))$ is in the relation. 
We show that

\BIT
There exist no correlation intractable function ensembles. 
\label{fail1.thm}
\EIT

\paragraph{Restricted correlation intractability.} 
The proof of the above negative result relies on the fact that 
the description of the function is shorter than its input. Thus 
we also investigate the case where one restricts the function $f_s$ 
to inputs whose length is less than the length of $s$. We show that 
the negative result can be extended to the case where the function 
description is shorter than the sum of the 
lengths of the input and output of the function. (Furthermore, if one 
generalizes the notion of \corInty\ to relations on sequences of inputs 
and outputs, then the negative result holds as long as the total length 
of all the inputs and outputs is more than the length of the function 
description.) 
This still leaves open the possibility that there exist function 
ensembles that are correlation intractable with respect to 
input-output sequences of a-priori bounded total length.  
See further discussion in Section~\ref{sec-lengths}.

\subsubsection{Failure of the Random Oracle Methodology}

Upon formulating the random oracle methodology, 
Bellare and Rogaway did warn 
that a proof of security in the Random Oracle Model should {\sc not} 
be taken as guarantee to the security of implementations (in which the 
Random Oracle is replaced by functions such as MD5)~\cite{BeRo93}. 
However, it is widely believed that a security proof in the {\ROM} 
means that there are no ``structural flaws'' in the scheme.
That is, any attack against an implementation of this scheme must 
take advantage of specific flaws in the function that is used to 
implement the oracle. In this work we demonstrate that these beliefs 
are false. Specifically, we show that 

\BIT
There exists encryption and signature schemes that are secure in the 
Random Oracle Model, but have {\sc no secure implementation} in the 
real model (where a Random Oracle does not exist).
That is, implementing these secure ideal schemes, using any function 
ensemble, results in insecure schemes. 
\label{fail2.thm}
\EIT
The encryption and signature schemes presented to prove 
Theorem~\ref{fail2.thm} are ``unnatural''.  We do not claim (or 
even suggest) that a statement as above holds with respect to 
schemes presented in the literature. 
Still, the lesson is that the mere fact that a scheme is secure 
in the Random Oracle Model does not necessarily imply that a particular 
implementation of it (in the real world) is secure, or even that this 
scheme does not have any ``structural flaws''. Furthermore, unless 
otherwise justified, such ideal scheme may have {\em no secure 
implementations at all}. 

In fact, our techniques are quite general and can be applied to
practically {\em any} cryptographic application. That is, given an 
ideal cryptographic application $A$, we can construct an 
ideal cryptographic application $A'$ such that 
$A'$ is just as secure as $A$ (in the \ROM), but $A'$ 
has {\em no secure implementation.} 
Hence, in this sense, security of an ideal system in the \ROM\ is a bad 
predictor of the security of an implementation of the system in real
life.

\subsection{Techniques}
Our proof of Theorem~\ref{fail2.thm} uses 
in an essential way 
non-interactive CS-proofs (in the \ROM),
as defined and constructed by Micali~\cite{Mi94}.%  
\footnote{\ 
The underlying ideas of Micali's construction~\cite{Mi94} 
can be traced to Kilian's construction~\cite{Ki92}
and to the Fiat--Shamir transformation~\cite{FiSh86}
(which is sound in the \ROM).}  
Interestingly, we only use the fact that non-interactive CS-proofs 
exist in the \ROM, and do not care whether or not these ideal CS-proofs 
have an implementation using any function ensembles 
(nor if non-interactive CS-proofs exists at all outside of the \ROM).
Specifically, CS-proofs are used to ``effectively verify''  
{\em any}\/ polynomial-time verifiable statement within time that 
is bounded by one {\em fixed}\/ polynomial. 
Furthermore, we use the fact that the definition of
CS-proofs guarantees that the complexity of generating such proofs 
is polynomial in the time required for ordinary verification.
See further discussion in Section~\ref{sec:csproofs}.

\subsection{Related Work}
%------------------------

\paragraph{Correlation intractability.} 
Our definition of correlation-intractability is related to a definition 
by Okamoto \cite{Ok92}. Using our terminology, Okamoto considers 
function ensembles for which it is infeasible to form input-output 
relations with respect to a specific evasive relation~\cite[Def.~19]{Ok92} 
(rather than all such relations).
He uses the assumption that such function ensembles exists,
for a specific evasive relation in \cite[Thm.~20]{Ok92}.

\paragraph{Special-purpose properties of the \ROM.} 
First steps in the direction of identifying and studying useful 
special-purpose properties of the \ROM\ have been taken by 
Canetti~\cite{Ca97}. Specifically, Canetti considered a property 
called ``perfect one-wayness'', provided a definition of this 
property, constructions which possess this property (under some 
reasonable assumptions), and applications for which such functions 
suffice. Additional constructions have been suggested by 
Canetti, Micciancio and Reingold~\cite{CMR98}. Another context where
specific properties of the random oracle where captured and realized
is the signature scheme of Gennaro, Halevi and Rabin \cite{GHR99}.

\paragraph{Relation to Zero-Knowledge proofs.} 
Following the preliminary version of the current work~\cite{CGH98}, 
Hada and Tanaka 
observed that the existence of even restricted correlation intractable 
functions (in the non uniform model) would be enough to prove that 
3-round auxiliary-input zero-knowledge AM proof systems only exist 
for languages in BPP \cite{HaTa99}. 
(Recall that auxiliary-input
zero-knowledge is seemingly weaker than black-box zero-knowledge,
and so the result of~\cite{HaTa99} is incomparable to prior work
of Goldreich and Krawczyk~\cite{GK} that showed 
that constant-round auxiliary-input zero-knowledge
AM proof systems only exist for languages in BPP.) 

\paragraph{Relation to ``magic functions''.} 
More recently, Dwork~{\etal} investigated  the notion 
of ``magic functions'', which is related to our {\corInt} 
functions \cite{magic}.  
Like {\corInty}, the definition of ``magic functions'' is 
motivated by the quest to capture the properties that are required from 
the hash function in the Fiat-Shamir heuristic. 
Correlation intractability seems like a general and natural property,
but is not known to be either necessary or sufficient for the 
Fiat-Shamir heuristic (which is a special case of the random oracle
methodology). In contrast, ``magic functions'' are explicitly 
defined as ``functions that make the Fiat-Shamir heuristic work''.  
In their paper \cite{magic}, Dwork~{\etal} demonstrated a relation 
between ``magic functions'' and 3-round zero-knowledge, 
which is similar to the relation between {\corInty} 
and zero-knowledge exhibited in \cite{HaTa99}. 
Specifically, they showed that the existence of ``magic functions'' 
implies the non-existence of some kind of 3-round zero-knowledge proof 
systems, as well as a weakened version of a converse theorem.

\subsection{Organization}
\secref{sec-syntax} presents syntax necessary for the rest of 
the paper as well as review the definition of CS-proofs. 
\secref{sec-unpredict} discusses the reasoning that 
led us to define the {\corInty} property, and prove that even such 
a minimalistic definition cannot be met by a function ensemble. 
\secref{sec-insecure} presents our main negative results -- 
demonstrating the existence of secure ideal signature and encryption 
schemes that do not have secure implementations. Restricted {\corInty} 
is defined and studied in \secref{sec-lengths}. 
Three different perspectives on the results obtained in this paper are
presented in \secref{sec-conclusions}.

%%%%%%%%%%%%%%%%%%%%%%%%%%%%%%%%%%%%%%%%%%%%%%%%%%%%%%%%%%%%%%
\section{Preliminaries}\label{sec-syntax}
%%%%%%%%%%%%%%%%%%%%%%%%%%%%%%%%%%%%%%%%%%%%%%%%%%%%%%%%%%%%%%
We consider probability spaces defined over executions
of probabilistic machines. Typically, we consider the probability
that an output generated by one machine $M_1$ satisfies a
condition that involves the execution of a second machine $M_2$.
For example, we denote 
by $\prob[y\gets M_1(x)\,,\,|y|\!=\!|x|\,\&\,M_2(y)\!=\!1]$
the probability that on input $x$, machine $M_1$ outputs 
a string that has length $|x|$ and is accepted by machine $M_2$. 
That is, $y$ in the above notation represents a random variable
that may be assigned arbitrary values in $\bitset^*$,
conditions are made regarding this $y$,
and we consider the probability that these conditions are
satisfied when $y$ is distributed according to $M_1(x)$. 

\subsection{Function Ensembles}\label{sec-ensembles}
To make the discussion in the Introduction more precise, we explicitly 
associate a length function, $\rlen:\intfnc$, with the output of the 
random oracle 
and its candidate implementations. We always assume that the length 
functions are super-logarithmic and polynomially bounded (i.e. 
$\omega(\log k) \leq \rlen(k) \leq \poly(k)$).  We refer to an 
oracle with length function $\rlen$ as an {\sf $\rlen$-oracle}. 
On security parameter $k$, each answer of the oracle  is a 
string of length $\rlen(k)$. A candidate implementation of a random 
$\rlen$-oracle is an $\rlen$-ensemble as defined below.

\BD[function ensembles] \label{def-ensemble}
Let $\rlen : \intfnc$ be a length function. 
An {\sf $\rlen$-ensemble} 
is a sequence ${\cal F} = \{F_k\}_{k \in N}$ of families of functions, 
$F_k = \{ f_s:\bitset^*\!\to\!\bitset^{\rlen(k)}\}_{s \in \bitset^k}$, 
so that the following holds
\begin{description}
\item[{\rm Length requirement.}]  
For every $s\in\bitset^k$ and every $x\in\bitset^*$, $|f_s(x)| = \rlen(k)$. 
\item[{\rm Efficiency requirement.}]
There exists a polynomial-time algorithm {\sc Eval} 
so that for every $s, x \in \bitset^*$, 
it holds that $\mbox{\sc Eval}(s,x) = f_s(x)$.
\end{description}
In the sequel we often call $s$ the {\sf description} 
or the {\sf seed} of the function $f_s$. 
\ED

\BRK \label{rmk-seed-len}
The length of the seed in the above definition serves as 
a ``security parameter'' and is meant to control the ``quality'' 
of the implementation. 
It is important to note that although $f_s(\cdot)$ is syntactically 
defined on every input, in a cryptographic applications it is only used 
on inputs of length at most $\poly(|s|)$. We stress that all results 
presented in this paper refer to such usage. 
\ERK

\BRK \label{rmk-restricted-corint}
One may even envision applications in which a more stringent condition 
on the use of $f_s$ holds. Specifically, one may require that the 
function $f_s$ be only applied to inputs of length at most $\dlen(|s|)$, 
where $\dlen : \intfnc$ is a specific (polynomially bounded) length 
function (e.g., $\dlen(k) = 2k$).  We discuss the effects of making 
such a stringent requirement in \secref{sec-lengths}. 
\ERK

\subsection{CS Proofs} \label{sec:csproofs}
\def\prv{\mbox{\sc Prv}}
\def\badP{\mbox{\sc Bad}}
\def\ver{\mbox{\sc Ver}}

Our construction of signature and encryption schemes that are secure in 
the {\ROM} but not in the ``real world'' uses CS-proofs as defined and 
constructed by Micali~\cite{Mi94}. 
Below, we briefly recall the relevant definitions and results.

A CS-proof system consists of a prover, $\prv$, 
that is trying to convince a verifier, $\ver$, 
of the validity of an assertion of the type 
{\em machine $M$ accepts input $x$ within $t$ steps}.\footnote{\
When $t$ is presented in binary, such valid assertions form
a complete language for the class (deterministic) exponential time.} 
The central feature of CS-proofs is that the running-time of the prover 
on input $x$ is (polynomially) related to the {\em actual} running time 
of $M(x)$ rather than to the global upper bound $t$; furthermore, the
verifier's running-time is poly-logarithmic related to $t$. 
(These conditions are expressed in the {\em additional efficiency 
requirements} in \defref{def-cs-proofs} below.)

In our context, we use non-interactive CS-proofs 
that work in the \ROM; that is, 
both prover and verifier have access to a common random oracle.
The prover generates an alleged proof 
that is examined by the verifier. 
A construction for such CS-proofs was presented by Micali~\cite{Mi94}, 
using ideas that can be traced to Kilian's construction~\cite{Ki92},
and requires no computational assumptions. 
Following is the formulation of CS-proofs, as defined in~\cite{Mi94}.

In the formulation below, the security parameter $k$ 
is presented in unary to both parties, whereas the global time bound 
$t$ is presented in unary to the prover and in binary to the verifier. 
This allows the (polynomial-time) prover to run in time polynomial
in $t$, whereas the (polynomial-time) verifier may only run in
time that is poly-logarithmic in $t$. 
(Observe that it is \emph{not required} that $t$ is bounded above by 
a polynomial in $|x|$. In fact, in our arguments, we shall use a 
slightly super-polynomial function $t$ (i.e., $t(n)=n^{\log n}$).) 
Finally, we mention that both the prover and the verifier in the definition 
below are required to be deterministic machines. See some discussion in 
Remark~\ref{rmk-det-verifier} below.

\BD[Non-interactive CS proofs in the \ROM] \label{def-cs-proofs}
A {\sf CS-proof system} consists of two {\em(deterministic)} 
polynomial-time oracle machines, a prover $\prv$ and a verifier 
$\ver$, which operate as follows: 
\begin{itemize}
\item 
On input $(1^k,\vect{M},x,1^t)$ and access to an oracle $\f$, 
the prover computes a proof $\pi = \prv^\f(1^k,\vect{M},x,1^t)$
such that $|\pi|=\poly(k,|\vect{M}|,|x|,\log t)$. 

\item 
On input $(1^k,\vect{M},x,t,\pi)$, with $t$ encoded in binary, 
and access to $\f$, 
the verifier decides whether to accept or reject the proof $\pi$
{\rm(i.e., $\ver^\f(1^k,\vect{M},x,t,\pi)
               \in\{{\tt accept},{\tt reject}\}$)}.
\end{itemize}
The proof system satisfies the following conditions, where the 
probabilities are taken over the random choice of the oracle $\f$:

\begin{description}
\item[{\rm Perfect completeness:}]
 For any $M,x,t$ such that machine $M$ accepts the string $x$ within
 $t$ steps, and for any $k$,
 \[\Pr_\f\left[
     \begin{array}{l}\pi \gets \prv^\f(1^k,\vect{M},x,1^t),\\
     \ver^\f(1^k,\vect{M},x,t,\pi)={\tt accept}\end{array}\right]
    \; = 1 \]

\item[{\rm Computational soundness:}]
 For any polynomial time oracle machine $\badP$ and any input $w =
 (\vect{M},x,1^t)$ such that $M$ {\em does not} accepts $x$ within
 $t$ steps, it holds that
 \[\Pr_\f\left[\begin{array}{l}
                \pi \gets \badP^\f(1^k,\vect{M},x,1^t),\\
                \ver^\f(1^k,\vect{M},x,t,\pi)={\tt accept}
               \end{array} \right] \leq \frac{poly(k+|w|)}{2^k} \]

\item[{\rm Additional efficiency conditions:}\footnotemark]
 The running-time of the prover $\prv$ on input $(1^k, \vect{M},x,1^t)$ 
 is (polynomially) related to the {\em actual} running time of $M(x)$, 
 rather than to the global upper bound $t$.  That is, there exists a 
 fixed polynomial $p(\cdot)$, such that 
 \[
   T_{_{PRV}}\left(1^k, \vect{M},x,1^t\right) \leq p(k,\min\{t,T_M(x)\})
 \] 
 where $T_A(x)$ denotes the running time of machine $A$ on input $x$. 
\footnotetext{\
By the above, the running time of $\prv$ on input $(1^k,\vect{M},x,1^t)$
is at most $\poly(k,|\vect{M}|,|x|,t)$, 
whereas the running time of $\ver$ on input $(1^k,\vect{M},x,t,\pi)$
is at most $\poly(k,|\vect{M}|,|x|,|\pi|,\log t)$.
The following condition provide even lower running time bound
for the prover.} 
\end{description}
\ED

\BR{Oracle output length} \label{rmk-output-length}
The above definition does not specify the output length of the oracle 
(i.e., the length of the answers to the oracle queries). In some cases 
it is convenient to identify this output length with the security 
parameter, but in many case we do not follow this convention (e.g., 
in \prpref{prp-cs-prf} below). 
In any case, it is trivial to implement an oracle with one output 
length given an oracle with different output length, so we allow 
ourselves to ignore this issue. 
\ER

\BR{Deterministic verifier} \label{rmk-det-verifier}
Recall that \defref{def-cs-proofs} mandates that both the prover and 
verifier are deterministic. 
Indeed this deviates from the tradition (in this area) of
allowing the verifier to be probabilistic; but Micali's
construction (in the \ROM) happens to employ a deterministic
verifier (cf.~\cite{Mi94}). This issue is not essential to our
main results, but plays an important role in the proof of
Proposition~\ref{Nissim.prop2} (due to K.~Nissim).
We note that when working in the \ROM\/
(and only caring about completeness and soundness),
one may assume without loss of generality that the prover is
deterministic (because it can obtain adequate randomness by
querying the oracle). This does not hold with respect to the
verifier, since its coin tosses must be unknown to the prover. 
\ER

\BT[Micali~\cite{Mi94}] \label{thm-cs-proofs}
There exists a non-interactive CS proof system in the \ROM. 
\ET 

For the proof of our construction (\thmref{sign.thm}), we will 
need a different soundness condition than the one from above.
Specifically, we need to make sure that given the machine $M$
(and the complexity bound $t$), it is hard to find {\em any pair}
$(x,\pi)$ such that $M$ does not accept $x$ within $t$ steps and yet
$\ver$ will accept $\pi$ as a valid CS-proof to the contrary.
One way to obtain this soundness property from the original one,
is by postulating that when the verifier is given a proof for an
assertion $w = (\vect{M},x,t)$, 
it uses security parameter $k+|w|$ (rather than just $k$).  
Using a straightforward counting argument we get: 

\BP{~}\label{prp-cs-prf} 
Let $(\prv,\ver)$ be a CS proof system. % in the \ROM. 
Then for every polynomial time 
oracle machine $\badP$, there exists a polynomial $q(\cdot)$, such that 
for every $k$ it holds that
\[
\e_{\rm bad}(k) \eqdef
\Pr_\f\left[\begin{array}{l}
 (\pi,w) \gets \badP^\f(1^{k}), \mbox{ where } w = (\vect{M},x,t),\\
 \mbox{s.t. machine $M$ does not accept $x$ within $t$ steps}\\
 \ \ \ \ \mbox{ and yet }\ver^\f(1^{k+|w|},w,\pi)={\tt accept}
\end{array} \right] \leq \frac{q(k)}{2^{k}}
\]
\EP

%%%%%%%%%%%%%%%%%%%%%%%%%%%%%%%%%%%%%%%%%%%%%%%%%%%%%%%%%%%%%%
\section{Correlation Intractability} \label{sec-unpredict}
%%%%%%%%%%%%%%%%%%%%%%%%%%%%%%%%%%%%%%%%%%%%%%%%%%%%%%%%%%%%%%
In this section we present and discuss the difficulty 
of defining the intuitive requirement that a function ensemble 
``behaves like a random oracle'' even when its description is given. 
In particular, we show that
even some minimalistic definitions cannot be realized.

\paragraph{An obvious failure.}
We first comment that an obvious maximalistic definition, which amount 
to adopting the pseudorandom requirement of~\cite{ggm86}, fails poorly. 
That is, we cannot require that an (efficient) algorithm that is
given the description of the function cannot distinguish its input-output 
behavior from the one of a random function, because the function 
description determines its input-output behavior. 

\paragraph{Towards a minimalistic definition.}
Although we cannot require the value of a fully specified function 
to be ``random'', we may still be able to require that it has some 
``unpredictability properties''.  
For example, we may require that, given a description 
of a family and a function chosen at random from a this family,
it is hard to find two preimages that the function maps to the same 
image. Indeed, this sound definition coincides with the well-known 
{\em collision-intractability} property \cite{Da87}. 
Trying to generalize, we may replace the ``equality of images'' relation 
by any other relation among the pre-images and images of the function. 
Namely, we would like to say that an ensemble is {\em \corInt\ } if for 
{\em any} relation, given the description of a randomly chosen function, 
it is infeasible to find a sequence of preimages that together with their 
images satisfy this relation. 

This requirement, however, is still unreasonably strong since there 
are relations that are easy to satisfy even in the \ROM. We therefore 
restrict the above infeasibility requirement by saying that it holds 
only with respect to relations that are hard to satisfy in the \ROM. 
That is, {\sc if} it is hard to find a sequence of preimages that together 
with their images under a random function satisfy relation $R$, {\sc then} 
given the description of a randomly chosen function $f_s$ it should be 
hard to find a sequence of preimages that together with their images 
under $f_s$ satisfy $R$. 

This seems to be a minimalistic notion of \corInt\ ensemble of 
functions, yet we show below that no ensemble can satisfy it. 
In fact, in the definition below we only consider the task of 
finding a single preimage that together with its image satisfies some 
property. Namely, instead of considering all possible relations, we only 
consider binary ones. Since we are showing impossibility result, this 
syntactic restriction only strengthens the result.

\subsection{Actual Definitions} 

We start with a formal definition of a relation 
that is hard to satisfy in the random oracle model.

\BD[Evasive Relations]
\label{def-admissible}
A binary relation $R$ is said to be {\sf evasive} with respect 
to length function $\rlen$ if for any 
probabilistic polynomial time oracle machine $M$ 
$$\Pr_\f[x\gets M^\f(1^k),\;\, % \mbox{\rm s.t.}\;\;|x|=k \mbox{\rm  and }
	(x,\f(x))\!\in\!R]\;\;=\;\;\negl(k)$$
where $\f:\bitset^*\to\bitset^{\rlen(k)}$ is a uniformly 
chosen function and $\negl(\cdot)$ is a negligible function.%
\footnote{\ 
A function $\mu\!:\!\N\to\!R$ is negligible if for every positive 
polynomial $p$ and all sufficiently large $n$'s, $\mu(n)<1/p(n)$.
} 
\ED
A special case of evasive relations consists of $R$'s for which there 
exists a negligible function $\negl(\cdot)$ so that for all $k$ 
$$\max_{x \in \bitset^*}\ \ \left\{
  \Pr_{y\in\bitset^{\rlen(k)}}[(x,y)\!\in\!R\,]\,\right\}\ =\ \negl(k)
$$
(All the binary relations used in the sequel falls into this category.) 
The reason such an $R$ is evasive is that any oracle machine, $M$, 
making at most $\poly(k)$ queries to a random $\f$ satisfies
\begin{eqnarray*}
\Pr_\f[x\gets M^\f(1^k),\;\, (x,\f(x))\!\in\!R] 
&\leq& \poly(k)\cdot
	\max_{x \in \bitset^*}\{\ \Pr_\f[(x,\f(x))\!\in\!R]\ \}\\
&\leq& \poly(k) \cdot \negl(k) 
\end{eqnarray*}

We are now ready to state our minimalistic definition 
of a \corInt\ ensemble: 
\BD[\CorInty]
\label{def-unpredict}
Let $\rlen : N \to N$ be length function, 
and let ${\cal F}$ be an $\rlen$-ensemble. 
\BI
\item 
Let $R\subseteq\bitset^*\times\bitset^*$ be a binary relation. 
We say that ${\cal F}$ is {\sf \corInt\/ with respect to $R$} 
if for every probabilistic polynomial-time machine $M$ 
it holds that
\[
\Pr_{s \in \bitset^k}[ x \gets M(s),\;\; %|x| = k \mbox{\rm  and }
	(x,f_s(x)) \in R ] = \negl(k)
\]
where $\negl(\cdot)$ is a negligible function, and the probability 
is taken over the choice of $s \in \bitset^k$ and the coins of $M$. 
\item 
We say that ${\cal F}$ is {\sf \corInt} if it is \corInt\/ with
respect to every evasive (w.r.t. $\rlen$) relation, 
\EI 
\ED

\BRK \label{rmk-weak-corint}
In the above definition we quantify over all evasive relations. 
A weaker notion, called {\sf weak \corInty}, is obtained by quantifying 
only over all polynomial-time recognizable evasive relations (i.e., 
we only consider those relations $R$ such that there exists a polynomial 
time algorithm that, given $(x,y)$, decides whether or not $(x,y) \in R$). 
In the sequel we consider both notions. 
\ERK

\subsection{Correlation-intractable ensembles do not exist} 
\label{sec.no-corint}
%----------------------------------------------------------

\BT\label{no-impl1} 
There exist no \corInt\ ensembles, not even in the weak sense. 
\ET

\begin{proof}
Let $\rlen$ be a length function and let ${\cal F}=\{f_s\}$
be an $\rlen$-ensemble. We define the binary relation: 

\begin{equation}
\label{eq-rf}
R^{\cal F} \eqdef \bigcup_k \left\{(s,f_s(s)) : s\in\bitset^k\right\}
\end{equation}
Clearly, this relation is polynomial-time recognizable, since $f_s$ 
can be computed in polynomial time. Also, the relation is evasive 
(w.r.t. $\rlen$) since for every $x \in \bitset^*$ there is at most
one $y \in \bitset^{\rlen(k)}$ satisfying $(x,y) \in R^{\cal F}$, 
\footnote{
Such a $y$ exists if and only if $\rlen(|x|)=\rlen(k)$.
}\ 
and so 
$$\Pr_y[(x,y) \in R^{\cal F}] \leq 2^{-\rlen(k)}=2^{-\omega(\log k)}
=\negl(k)\,.
$$ 
On the other hand, consider the machine $I$ that computes the 
identity function, $I(x) = x$ for all $x$. It violates the \corInty\ 
requirement, since for all $k$, 
$$
\Pr_{s\in\bitset^k}[(I(s),f_s(I(s)))\in R^{\cal F}]
\ =\ \Pr_{s\in\bitset^k}[(s,f_s(s))\in R^{\cal F}]
\ =\ 1\,.
$$
In fact, since $R^{\cal F}$ is polynomial-time recognizable,
even the weak \corInty\ of $\cal F$ is violated. 
\end{proof}

%%%%%%%%%%%%%%%%%%%%%%%%%%%%%%%%%%%%%%%%%%%%%%%%%%%%%%%%%%%%%%%%%%%
\section{Failure of the Random Oracle Methodology}
\label{sec-insecure}
%%%%%%%%%%%%%%%%%%%%%%%%%%%%%%%%%%%%%%%%%%%%%%%%%%%%%%%%%%%%%%%%%%%
This section demonstrates that the security of a cryptographic scheme 
in the \ROM\/ does not always imply its security under some specific 
choice of a ``good hash function'' that is used to implement 
the random oracle.  To prove this statement we construct signature and 
encryption schemes, which are secure in the \ROM, yet for which {\em 
any implementation} of the random oracle yield insecure schemes. Put 
in other words, although the ideal scheme is secure, any implementation 
of it is necessarily insecure. 

The underlying idea is to start with a secure scheme (which may or may 
not use a random oracle) and modify it to get a scheme that is secure 
in the \ROM, but such that its security is easily violated when trying 
to replace the random oracle by any ensemble. This is done by using  
evasive relations as constructed in \thmref{no-impl1}. 
The modified scheme starts by trying to find a preimage that 
together with its image yields a pair in the evasive relation. In case 
the attempt succeeds, the scheme does something that is clearly insecure 
(e.g., 
output the secret key). 
Otherwise, the scheme behaves as the original (secure) 
scheme does. The former case (i.e., finding a pair in the relation) will 
occur rarely in the \ROM, thus the scheme will maintain its security 
there. However, it will be easy for an adversary to make sure that the 
former case always occurs under any implementation 
of the \ROM,  thus no implementation may be secure.  
We start with the case of a signature scheme, 
and present the construction in three steps. 

\BI
\item
In the first step we 
carry out the above idea in a naive way. This allows us to
prove a weaker statement, saying that for any function ensemble 
$\cal F$, there exists a signature scheme that is secure in the 
\ROM, but is not secure when implemented using $\cal F$. 

This, by itself, means that one cannot construct a function ensemble 
that provides secure implementation of any cryptographic scheme that 
is secure in the \ROM.  But it does not rule out the possibility (ruled 
out below) that for any cryptographic scheme that is secure in the 
\ROM\ there exists a secure implementation (via a different function 
ensemble). 

\item 
In the second step we use diagonalization techniques to reverse the 
order of quantifiers. Namely, we show that there exists a signature 
scheme that is secure in the \ROM, but for which {\em any} implementation
(using any function ensemble) results in an insecure scheme. However, 
the scheme constructed in this step utilizes signing and verification 
procedures that run in (slightly) super-polynomial time.

\item 
In the third step we use CS-proofs \cite{Mi94} to get rid of the 
super-polynomial running-time (of the legitimate procedures), 
hence obtaining a standard signature scheme that is secure in the \ROM, 
but has no secure implementation. 
Specifically, in this step we use CS-proofs as a tool to 
``diagonalize against all polynomial-time ensembles in polynomial 
time''. (As noted by Silvio Micali, this technique may be useful also 
in other settings where diagonalization techniques are applied.) 
\EI
The reader is referred to~\cite{gmr88} for basic terminology regarding
signature schemes and corresponding notions of security. As a starting 
point for our constructions, we use a signature scheme, denoted 
${\cal S} = (G,S,V)$, where $G$ is the key-generation algorithm, $S$ is 
the signing algorithm, and $V$ is the verification algorithm. We assume 
that the scheme $(G,S,V)$ is existentially unforgeable under adaptive 
chosen message attack, in the \ROM.
We do not need to rely on any computational 
assumptions here, since one-way functions are sufficient for
constructing secure signature schemes~\cite{ny89,ro90}, and the random
oracle can be used to implement one-way functions without any
assumptions.%
\footnote{\ Alternatively, we could use an `ordinary' signature scheme, 
but then our Theorem~\ref{sign.thm} would be conditioned on the existence 
of one-way functions.} 

\paragraph{Conventions.}
In the three steps below we assume, without loss of generality, that 
the security parameter (i.e., $k$) is implicit in the keys generated 
by $G(1^k)$. Also, let us fix some length function $\rlen : \intfnc$, 
which would be implicit in the discussions below (i.e., we assume that 
the random oracles are all $\rlen$-oracles, the relations are evasive 
w.r.t.~$\rlen$, etc.).

\subsection{First Step}\label{step1.sec}
%---------------------------------------
\paragraph{Definition.} 
{\em Let ${\cal S} = (G,S,V)$ be a signature scheme 
(which may or may not use a random oracle), and let $R$ be any binary 
relation, which is evasive w.r.t.~length function $\rlen$. 
Then, by ${\cal S}_R = (G,S_R,V_R)$ we denote the following modification 
of ${\cal S}$ which utilizes a random $\rlen$-oracle:} 
\begin{description}
\item[{\rm Modified signature, $S_R^\f(\sk,\msg)$,}]  
{\em of message $\msg$ using signing key $\sk$:

 1. If $(\msg,\f(\msg))\!\in\!R$, output $(\sk,\msg)$. 

 2. Otherwise (i.e., $(\msg,\f(\msg))\!\not\in\!R$),
    output $S^{\f}(\sk,\msg)$.}

\item[{\rm Modified verification, $V_R^\f(\vk,\msg,\sigma)$,}]   
{\em of alleged signature $\sigma$ to $\msg$ using verification key $\vk$:

 1. If $(\msg,\f(\msg))\!\in\!R$ then {\tt accept}

 2. Otherwise output $V{^\f}(\vk,\msg,\sigma)$.} 
\end{description}
The key-generation algorithm, $G$, is the same as in the 
original scheme ${\cal S}$.  Item~1 in the signing/verification 
algorithms is a harmful modification to the original signature scheme. 
Yet, if $R$ is evasive, then it has little effect on the ideal system, 
and the behavior of the modified scheme is ``indistinguishable'' from 
the original one. In particular, 

\BP\label{signROM} 
Suppose that $R$ is evasive {\em(w.r.t.~$\rlen$)} and that ${\cal S}$ 
is existentially unforgeable under a chosen message attack in the \ROM. 
Then ${\cal S}_R$ is also existentially unforgeable under a chosen message 
attack in the \ROM. 
\EP 

\begin{proof}
The intuition is that since $R$ is evasive, it is infeasible for the 
forger to find a message $m$ so that $(m,\f(m))\in R$. Thus, a forgery 
of the modified scheme must be due to Item~(2), which yields a breaking 
of the original scheme. 

Formally, let $A_R$ be an adversary who mounts an adaptive chosen 
message attack on ${\cal S}_R$, and whose success probability in 
obtaining an existential forgery (in the \ROM) is $\epsilon_{\rm frg} 
= \epsilon_{\rm frg}(k)$. Assume, toward contradiction, that 
$\epsilon_{\rm frg}$ is not negligible in the security parameter $k$.

Denote by REL the event in which during an execution of $A_R$, it 
hands out a message $m$ for which $(m,\f(m))\!\in\!R$ (either as a query 
to the signer during the chosen message attack, or as the message for 
which it found a forgery at the end), and let $\epsilon_{\rm rel} = 
\epsilon_{\rm rel}(k)$ be the probability of that event. 
Using the hypothesis that $R$ is evasive, 
we prove that $\epsilon_{\rm rel}$ is negligible 
in the security parameter $k$. 
Suppose, to the contrary, that $\epsilon_{\rm rel}$ is not negligible.
Then, we can try to efficiently find pairs $(x,\f(x))\in R$ by 
choosing a key-pair for $\cal S$, and then implementing the attack, 
playing the role of both the signer algorithm and the adversary $A_R$. 
With probability $\epsilon_{\rm rel}$, one of $A_R$'s messages during 
this attack satisfies $(m,\f(m))\!\in\!R$, so just choosing at random 
one message that was used and outputting it yields a success probability 
of $\epsilon_{\rm rel}/q$ (with $q$ being the number of different 
messages that are used in the attack). If $\epsilon_{\rm rel}$ is not 
negligible, then neither is $\epsilon_{\rm rel}/q$, contradicting 
the evasiveness of $R$. 

It is clear that barring the event REL, the execution of $A_R$ 
against the original scheme ${\cal S}$ would be identical to its 
execution against ${\cal S}_R$. 
Hence the probability that $A_R$ succeeds in obtaining an existential 
forgery against ${\cal S}$ is at least $\epsilon_{\rm frg} - 
\epsilon_{\rm rel}$. Since $\epsilon_{\rm rel}$ is negligible, 
and $\epsilon_{\rm frg}$ is not, then $A_R$'s probability of obtaining 
an existential forgery against ${\cal S}$ is also not negligible, 
contradicting the assumed security of ${\cal S}$. 
\end{proof}

\bigskip\ni
The modification enables to break the modified scheme when implemented 
with a real ensemble $\cal F$, in the case where $R$ is the relation 
$R^{\cal F}$ from Proposition~\ref{no-impl1}.  Indeed, as corollary 
to Propositions~\ref{no-impl1} and~\ref{signROM}, we immediately obtain:

\BCR\label{weak.cor} 
For every efficiently computable $\rlen$-ensemble ${\cal F}$, 
there exists 
a signature scheme that 
is existentially unforgeable under a chosen message attack 
in the \ROM,
yet when 
implemented with $\cal F$, the resulting scheme is totally breakable 
under an adaptive chosen message attack, and existentially forgeable 
under a key-only attack. 
\ECR

\begin{proof}
When we use an ensemble ${\cal F}$ to implement the random 
oracle in the scheme ${\cal S}_R$, we obtain the following real scheme 
(which we denote ${\cal S}'_R = (G', S'_R, V'_R)$): 
\begin{description}
\item[$G'(1^k)$: ] Uniformly pick $s \in \bitset^k$, set $(\sk$,
 $\vk)$ $\gets G^{f_s}(1^k)$, and output $(\vect{\sk,s},\vect{\vk,s})$. 

\item[$S'_R(\vect{\sk,s},\msg)$:] 
Output $S_R^{f_s}(\sk,\msg)$. 

\item[$V'_R(\vect{\vk,s},\msg,\sigma)$:] 
Output $V_R^{f_s}(\vk,\msg,\sigma)$. 
\end{description}
Consider now what happens when we use the ensemble ${\cal F}$ to 
implement the the scheme ${\cal S}_{R^{\cal F}}$ (recall the definition 
of $R^{\cal F}$ from \eqref{eq-rf}). Since $R^{\cal F}$ is evasive, then 
{from} Proposition~\ref{signROM} we infer that the ${\cal S}_{R^{\cal F}}$ 
is secure in the \ROM. 
However, when we use the ensemble ${\cal F}$ to implement the scheme, 
the seed $s$ becomes part of the public verification-key, and hence is known 
to the adversary. The adversary can simply output the pair $(s,\epsilon)$, 
which will be accepted by $V'_{R^{\cal F}}$ as a valid message-signature 
pair (since $(s, f_s(s)) \in R^{\cal F}$). Hence, the adversary achieves 
existential forgery (of ${\cal S}'_{R^{\cal F}}$) under key-only attack. 
Alternatively, the adversary can ask the legitimate signer for a signature 
on $s$, hence obtaining the secret signing-key (i.e., total forgery). 
\end{proof}

\subsection{Second Step}\label{step2.sec}
%------------------------
\paragraph{Enumeration.}
For this (and the next) subsection we need an enumeration of all 
efficiently computable function ensembles. Such enumeration is achieved 
via an enumeration of all polynomial-time algorithms (i.e., candidates 
for evaluation algorithms of such ensembles). 
Several standard technicalities arise. 
First, enumerating all polynomial-time algorithms is problematic since 
there is no single polynomial that bounds the running time of all these 
algorithms. Instead, we fix an arbitrary super-polynomial proper complexity 
function\footnotemark, $t:\N\!\to\!\N$ (e.g., $t(n) = n^{\log n}$), 
and enumerate all algorithms of running-time bounded by $t$. The latter 
is done by enumerating all possible algorithms, and modifying each 
algorithm by adding a time-out mechanism that terminates the execution 
in case more than $t(|{\rm input}|)$ steps are taken. This modification 
does not effect the polynomial-time algorithms. 
\footnotetext{\ Recall that $t(n)$ is a {\em proper complexity function} 
(or time-constructible) if there exists a machine that computes $t(n)$ 
and works in time $O(t(n))$. This technical requirement is needed to 
ensure that the enumeration itself is computable in time $O(t(n))$.}
Also, since we are interested in enumerating $\rlen$-ensembles, we 
modify each function by viewing its seed as a pair $\vect{s,x}$ 
(using some standard parsing rule\footnotemark) and padding or truncating 
its output to length $\rlen(|s|)$. Again, this modification has no 
effect on the $\rlen$-ensembles.
\footnotetext{\ For example, using a prefix-free code $C$, we can encode
a pair $(s,x)$ by $C(s)$ concatenated with $x$.}

Let us denote by ${\cal F}^i$ the $i^\th$ function ensemble according to 
the above enumeration, and denote by $f^i_s$ the function indexed by $s$ 
{from} the ensemble ${\cal F}^i$. Below we again use some standard 
rule for parsing a string $\alpha$ as a pair $\vect{i,s}$ and viewing 
it as a description of the function $f^i_s$.

\paragraph{Universal ensemble.} 
Let ${\cal U}=\{U_k\}_{k\in{\bf N}}$ 
denote the ``universal function ensemble'' that is induced 
by the enumeration above, namely
$U_k=\{u_{\vect{i,s}}\}_{\vect{i,s}\in\bitset^k}$ and 
$u_{\vect{i,s}}(x) = f^i_s(x)$.
There exists a machine that computes the universal 
ensemble $\cal U$ and works in slightly super-polynomial time, $t$.

\paragraph{Universal relation.}  
Denote by $R^{\cal U}$ the universal relation that is defined with 
respect to the universal ensemble $\cal U$ similarly to the way that 
$R^{\cal F}$ is defined with respect to any ensemble $\cal F$.
That is: 
\[
\label{eq-rf2}
R^{\cal U} \eqdef \bigcup_k \left\{\left(\vect{i,s},
f^i_s(\vect{i,s})\right) : \vect{i,s}\in\bitset^k\right\}
\]
Or, in other words:
\[ (x,y) \in R^{\cal U} \ \ \Longleftrightarrow
             \begin{array}{l}
             y =u_x(x) \\
             \mbox{\rm (i.e., $x = \vect{i,s}$ and $y = f^i_s(x)$)}
             \end{array}
\]

\paragraph{Modified signature scheme.} 
Let ${\cal S} = (G,S,V)$ be a signature scheme (as above). We then 
denote by ${\cal S}_u = (G,S_u,V_u)$ the modified signature scheme 
that is derived by using $R^{\cal U}$ in place of $R$ in the previous
construction. Specifically: 
\begin{description}
\item[$S_u^\f(\sk,\msg)$] $\eqdef$

1. If $(\msg, \f(msg)) \in R^{\cal U}$ (i.e., if $\msg=\vect{i,s}$ 
 and $\f(\msg)= f^i_s(\msg)$) then output $(\sk,\msg)$. 

2. Otherwise, output $S^\f(\sk,\msg)$

\item[$V_u^\f(\vk,\msg,\sigma)$] $\eqdef$ 

1. If $(\msg, \f(msg)) \in R^{\cal U}$ then {\tt accept}.

2. Otherwise, output $V^\f(\vk,\msg,\sigma)$.
\end{description}
We note that since these signature and verification algorithms need 
to compute $\cal U$, they both run in time $O(t)$, which is slightly 
super-polynomial. 

\BP \label{super-poly.prop}
Suppose that ${\cal S}$ is existentially unforgeable under a chosen 
message attack in the \ROM. Then ${\cal S}_u$ is also existentially 
unforgeable under a chosen message attack in the \ROM, but implementing 
it with {\em any function ensemble} yields a scheme 
that is totally breakable under chosen message attack 
and existentially forgeable under key-only attack.
\EP

\begin{proof}
Since $R^{\cal U}$ is evasive, then from \prpref{signROM} it follows 
that ${\cal S}_u$ is secure in the \ROM. On the other hand, suppose 
that one tries to replace the random oracle in the scheme by an ensemble 
${\cal F}^i$ (where $i$ be the index in the enumeration). An adversary, 
given a seed $s$ of a function in ${\cal F}^i$ can then set 
$\msg=\vect{i,s}$ 
and output the pair $(\msg,\epsilon)$, which would be accepted as a valid 
message-signature pair by $V_u$. Alternatively, it can ask the signer for 
a signature on this message $\msg$, and so obtain the secret signing-key. 
\end{proof}

\subsection{Third step}\label{step3.sec}
%----------------------
We now use CS-proofs to construct a new signature scheme that works in the 
\ROM. This construction is similar to the one in Subsection~\ref{step2.sec}, 
except that instead of checking that $(\msg,\f(\msg))\in R^{\cal U}$, 
the signer/verifier gets a CS-proof of that claim, and it only needs 
to verify the validity of that proof. Since verifying the validity of a 
CS-proof can be done much more efficiently than checking the claim ``from 
scratch'', the signing and verifications algorithms in the new scheme 
may work in polynomial time. On the other hand, when the scheme
is implemented using the function ensemble ${\cal F}^i$,
supplying the adequate CS-proof 
(i.e., for $(\msg,f_s^i(\msg))\in R^{\cal U}$)
only requires polynomial-time
(i.e., time polynomial in the time it takes to evaluate $f_s^i$).
This yields the following: 

\BT\label{sign.thm}
There exists a signature scheme that is existentially unforgeable under 
a chosen message attack in the \ROM, but such that when implemented with 
any function ensemble, the resulting scheme is existentially forgeable 
using key-only attack and totally breakable under chosen message attack. 
\ET
We note again that unlike the ``signature scheme'' presented in 
Subsection~\ref{step2.sec}, the signature scheme presented below works 
in polynomial-time. 
\medskip 

\begin{proof}
Below we describe such a signature scheme. 
For this construction we use the following ingredients.
\BI
\item ${\cal S} = (G,S,V)$ is a signature scheme, operating in the 
\ROM, that is existentially unforgeable under a chosen message attack.  

\item 
A fixed (and easily computable) parsing rule which interpret messages 
as triples of strings $\msg = \vect{i,s,\pi}$.

\item 
The algorithms $\prv$ and $\ver$ of a CS-proof system, as described 
in \secref{sec:csproofs} above. 

\item
Access to three independent random oracles. 
This is very easy to achieve given access to one oracle $\f$;
specifically, by setting $\f'(x) \eqdef \f(01x)$, $\f''(x) \eqdef \f(10x)$
and $\f'''(x) \eqdef \f(11x)$. 

Below we use oracle $\f'''$ for the basic scheme $\cal S$, oracle 
$\f''$ for the CS-proofs, and oracle $\f'$ for our evasive relation. 
We note that if $\f$ is an $\rlen$-oracle, then so are $\f', \f''$ 
and $\f'''$. 

\item 
The universal function ensemble 
${\cal U}$ from Subsection~\ref{step2.sec}, 
with proper complexity bound $t(n) = n^{\log n}$. 
We denote by $M_{\cal U}$ the universal machine that decides 
the relation $R^{\cal U}$.  
That is, on input $(\vect{i,s},y)$,
machine $M_{\cal U}$ invokes the $i^\th$ evaluation algorithm,
and accepts if $f^i_s(\vect{i,s}) = y$.

We note that $M_{\cal U}$ works in time $t$ in the worst case. 
More importantly, if ${\cal F}^i$ is a function ensemble that can be 
computed in time $p_i(\cdot)$ (where $p_i$ is some polynomial), then 
for any strings $s,y$, on input $(\vect{i,s},y)$, 
machine $M_{\cal U}$ works 
for only $\poly(|i|) \cdot p_i(|s|)$ many steps.%
\footnote{\ The point is merely that, for every fixed $i$, the 
expression $\poly(|i|) \cdot p_i(|s|)$ is bounded by a polynomial 
in $|s|$.} 
\EI
Using all the above, we describe an ideal signature scheme ${\cal S}'_u = 
(G,S'_u,V'_u)$.  As usual, the key generation algorithm, $G$, remains 
unchanged. The signature and verification algorithms proceed as follows. 

\begin{description}
\item[${S'_u}^\f(\sk,\msg)$] $\eqdef$  
\BE
\item
Parse $\msg$ as $\vect{i,s,\pi}$, and set $x=\vect{i,s}$ and 
$y=\f'(x)$. Let $n=|(x,y)|$. 
\item
Apply $\ver^{\f''}$ to verify whether $\pi$ is a valid CS-proof,
with respect to the oracle $\f''$ and security parameter $1^{n+k}$, 
for the claim that the machine $M_{\cal U}$ accepts the input $(x,y)$ 
within time $t(n)$. 

(The punch-line is that we do not directly check whether the
machine $M_{\cal U}$ accepts the input $(x,y)$ within time $t(n)$, 
but rather only if $\pi$ is a valid CS-proof of this claim.
Although $t(n)=n^{\log n}$, this CS-proof can be verified in
polynomial-time.) 
\item
If $\pi$ is a valid proof, then output $(\sk,\msg)$. 
\item
Otherwise, output $S^{\f'''}(\sk,\msg)$.
\EE

\item[${V'_u}^\f(\vk,\msg,\sigma)$] $\eqdef$  

{~~~}1+2. As above

{~~~}3. If $\pi$ is a valid proof, then {\tt accept}
 
{~~~}4. Otherwise, output $V^{\f'''}(\vk,\msg,\sigma)$.
\end{description}
The computation required in Item~2 of the signature 
and verification algorithms can be executed in polynomial-time. The 
reason being that (by definition) verifying a CS-proof can be done in 
polynomial-time, provided the statement can be decided in at most 
exponential time (which is the case here since we have $t(n) = 
O(n^{\log n})$). 
It is also easy to see that for every pair $(\sk,\vk)$ output by $G$, 
and for every $\msg$ and every $\f$, the string ${S'_u}^\f(\sk,\msg)$ 
constitutes a valid signature of $\msg$ relative to $\vk$ and the 
oracle $\f$. 

To show that the scheme is secure in the \ROM, we first observe that 
on security parameter $1^k$ it is infeasible to find a string $x$ so 
that $(x,\f'(x)) \in R^{\cal U}$, since $R^{\cal U}$ is evasive.
By \prpref{prp-cs-prf}, it is also infeasible to find $(x,\pi)$ 
such that $(x,\f'(x))\not\in R_{\cal U}$ and yet $\pi$ is a valid 
CS-proof of the contrary relative to $\f''$ (with security parameter 
$1^{|x|+\rlen(k)+k}$). 
Thus, it is infeasible for a polynomial-time adversary to find a message 
that would pass the test on Item~2 of the signature/verification algorithms 
above, and so we infer that the modified signature is secure in the \ROM. 

We now show that for every candidate implementation, ${\cal F}$, there 
exists a polynomial-time adversary effecting total break via a chosen 
message attack (or, analogously, an existential forgery via a ``key 
only'' attack). 
First, for each function $f_s \in {\cal F}$,
denote $f'_s(x) \eqdef f_s(01x)$, $f''_s(x) \eqdef f_s(10x)$, 
and $f'''_s(x) \eqdef f_s(11x)$. 
Then denote by ${\cal F}'$ the ensemble of the $f'_s$ functions. 

Suppose that ${\cal F}'$ is the $i^\th$ function ensemble in the 
enumeration mentioned above, namely ${\cal F}' = {\cal F}^i$. Given 
a randomly chosen $k$-bit seed $s$, the adversary generate a message 
$\msg=\vect{i,s,\pi}$ so that $\pi$ is a CS-proof (w.r.t the adequate 
security parameter) for the {\em true}\/ statement that $M_{\cal U}$ 
accepts the input $(x,y)$ within $t(|x|+|y|)$ steps, where $x = \vect{i,s}$ 
and $y = f'_s(x)$. Recall that the above statement is indeed true 
(since $f'_s\equiv f^i_s$),
and hence the adversary can generate a proof for it in time which is 
polynomial in the time that it takes to compute $f^i_s$. 
(By the perfect completeness property 
of the CS-proof system, the ability to prove 
correct statements holds for {\em any} choice of the random oracle, 
and in particular when it is equal to $f''_s$.) % See \AppendixCS.) 
Since this adversary is specifically designed to break 
the scheme in which the random oracle is implemented by ${\cal F}$, 
then the index $i$~-- 
which depends only on the choice of ${\cal F}$~-- 
can be incorporated into the program of this adversary. 

By the efficiency condition of CS-proofs, it is possible to find $\pi$ 
(given an oracle access to $f''_s$) in time polynomial in the time 
that it takes $M_{\cal U}$ to accept the input $(x,y)$. Since ${\cal F}^i$ 
is polynomial-time computable, then $M_{\cal U}$ works on the input $(x,y) 
= (\vect{i,s},y)$ in polynomial time, and thus the described adversary 
also operates in polynomial-time. 

By construction of the modified verification algorithm, $\epsilon$ is 
a valid signature on $\msg=\vect{i,s,\pi}$, and so existential forgery 
is feasible a-priori. Furthermore, requesting the signer to sign 
the message $\msg$ yields the signing key, and thus total forgery. 
\end{proof}

\BRK \label{rmk-csproofs-exist}
It is immaterial for the above argument whether CS-proofs 
can be implemented in the ``real world'' (i.e., without access to random 
oracles).  Specifically, it doesn't matter if one can cheat when the 
oracle is substituted by a candidate function ensemble, as in this case 
(i.e., in the real world implementation) it is sufficient for the adversary 
to invoke the proof system on valid statements. 
We do rely, however, on the perfect completeness of CS-proofs that 
implies that valid statements can be proven for any possible choice of 
oracle used in the proof system. 
\ERK

\subsection{Encryption}
%----------------------
The construction presented for signature schemes can be adapted to 
public-key 
encryption schemes in a straightforward 
way, yielding the following theorem:% 
\footnote{\ 
Similarly, we can adapt the argument to
shared-key (aka private-key) encryption schemes.
See Remark~\ref{rmk-similar-schemes}.}

\BT\label{encrypt.thm} 
{~} 
\begin{description}
\item[(a)] Assume that there exists a  public key encryption scheme that is 
semantically secure in the \ROM. Then there exists a public key 
encryption scheme that is semantically secure in the \ROM\ but is not 
semantically secure
when implemented with any function ensemble.% 
\footnote{Here we refer to semantic security
as defined by Goldwasser and Micali in~\cite{GoMi84}, and not to
the seemingly weaker definition presented in~\cite{G93,G99}.
Goldwasser and Micali allow the message space to depend on the
public-key, whereas this is not allowed in~\cite{G93,G99}.} 

\item[(b)] Assume that there exists a public key encryption scheme that is 
secure under adaptive chosen ciphertext attack in the \ROM. Then there 
exists a scheme that is secure under adaptive chosen ciphertext attack 
in the \ROM, but implementing it with any function ensemble yields a 
scheme that is not semantically secure, and in which a chosen ciphertext 
attack reveals the secret decryption key. 
\end{description}
\ET
\begin{proof}
In this proof we use the same notations as in the proof of \thmref{sign.thm}.
Let ${\cal E} = (G,E,D)$ be an encryption scheme that is semantically 
secure in the \ROM, and we modify it to get another scheme ${\cal E}' = 
(G,E',D')$. The key generation algorithm remains unchanged, and the 
encryption and decryption algorithms utilize a random oracle $\f$,
which is again viewed as three oracles $\f',\f''$ and $\f'''$. 
\begin{description}
\item[Modified encryption, ${E'_\ek}^\f(\msg)$,] of plaintext $\msg$
using the public encryption-key $\ek$:
\BE
\item Parse $\msg$ as $\vect{i,s,\pi}$, set $x = \vect{i,s}$ and 
$y = \f'(x)$, and let $n=|(x,y)|$.  

\item If $\pi$ is a valid CS-proof, w.r.t oracle $\f''$ and security 
parameter $1^{n+k}$, for the assertion that $M_{\cal U}$ accepts the 
pair $(x,y)$ within $t(n)$ steps, then output $(1,\msg)$. 

\item Otherwise (i.e., $\pi$ is not such a proof), output 
$(2,E_\ek^{\f'''}(\msg))$. 
\EE 

\item[Modified decryption, ${D'_\dk}^\f(c)$,] of ciphertext $c$
using the private decryption-key $\dk$:

\BE 
\item If $c = (1,c')$, output $c'$ and halt.

\item If $c = (2,c')$, output $D_{\dk}^{\f'''}(c')$ and halt.

\item If $c = (3,c')$ then parse $c'$ as $\vect{i,s,\pi}$, and set 
$x = \vect{i,s}$, $y = \f'(x)$, and $n=|(x,y)|$. 
If $\pi$ is a valid CS-proof, w.r.t oracle $\f''$ and security parameter 
$1^{n+k}$, for the assertion that $M_{\cal U}$ accepts the pair $(x,y)$ 
within $t(n)$ steps, then output $\dk$ and halt. 

\item Otherwise output $\epsilon$.
\EE 
\end{description}
The efficiency of this scheme follows as before. It is also easy to 
see that for every pair $(\ek,\dk)$ output by $G$, 
and for every plaintext $\msg$, 
the equality ${D'_\dk}^\f({E'_\ek}^\f(\msg))=\msg$ holds for every $\f$. 
To show that the scheme is secure in the \ROM, we observe again that it 
is infeasible to find a plaintext that satisfies the condition in Item~2 
of the encryption algorithm (resp., a ciphertext that satisfies the 
condition in Item~3 of the decryption algorithm). Thus, the modified 
ideal encryption scheme (in the \ROM) inherits all security features 
of the original scheme. 

Similarly, to show that replacing the random oracle by any function 
ensemble yields an insecure scheme, we again observe that for any 
such ensemble there exists an adversary who -- given the seed $s$ -- 
can generate a plaintext $\msg$ (resp., a ciphertext $c$) 
that satisfies the condition in Item~2 of the encryption algorithm 
(resp., the condition in Item~3 of the decryption algorithm). 
Hence, such an adversary can identify when $\msg$ is 
being encrypted (thus violates semantic security), or ask for a 
decryption of $c$, thus obtaining the secret decryption key. 
\end{proof}

\BRK \label{rmk-enc-exist}
As opposed to \thmref{sign.thm}, here we need 
to make computational assumptions, namely, that there exist schemes that 
are secure in the \ROM. 
(The result in~\cite{ImRu89} imply that it is unlikely that
such schemes are proven to exists without making any assumptions.)
Clearly, any scheme which is secure without random oracles is also secure 
in the \ROM. Recall that the former exist, provided trapdoor permutations 
exist~\cite{GoMi84,yao82a}. 
\ERK

\BRK \label{rmk-similar-schemes}
The constructions presented above can be adapted to yield
many analogous results.  For example, a result analogous to 
Theorem~\ref{encrypt.thm} holds for shared-key (aka private-key) 
encryption schemes.  In this case no computational assumptions 
are needed since secure shared-key encryption is known to exist 
in the \ROM.
Similarly, we can prove the existence of a CS-proof in the \ROM\/
that has no implementations (via any function ensemble). 
In fact,  as remarked in the Introduction,
the same technique can be applied to practically any cryptographic
application. 
\ERK

%----------------------------------------------
\section{Restricted correlation intractability}\label{sec-lengths}
%----------------------------------------------
Faced with the negative result of \thmref{no-impl1}, one may explore 
restricted (and yet possibly useful) versions of the \corInty\ property.
One possibility is to put more stringent constraints on the use of 
the ensemble in a cryptographic scheme, and then to show that as long 
as the ensemble is only used in this restricted manner, 
it is guaranteed to maintain some aspects of \corInty. 

In particular, notice that the proof of \thmref{no-impl1} relies heavily 
on the fact that the input to $f_s$ can be as long as the seed $s$, so 
we can let the input to the function $f_s$ be equal to $s$. 
Thus, one option 
would be to require that we only use $f_s$ on inputs that are shorter 
than $s$. 
Specifically, we require that each function $f_s$ will only be applied 
to inputs of length $\dlen(|s|)$, where $\dlen:\intfnc$ is some 
pre-specified function (e.g. $\dlen(k)=k/2$). The corresponding restricted 
notion of \corInty\ is derived from 
\defref{def-unpredict}: 

\BD[restricted \corInty]
\label{def-restrict-unpredict}
Let $\dlen,\rlen: \intfnc$ be length functions. A machine $M$ is called
$\dlen$-respecting if $|M(s)| = \dlen(|s|)$ for all $s\in\bitset^*$.
\BI 
\item 
A binary relation $R$ is evasive with respect to $(\dlen,\rlen)$ if for
any $\dlen$-respecting probabilistic polynomial-time oracle machine $M$
$$\Pr_\f[x\gets M^\f(1^k),\;\, (x,\f(x))\!\in\!R]\;\;=\;\;\negl(k)$$
where $\f:\bitset^{\dlen(k)}\to\bitset^{\rlen(k)}$ is a uniformly
chosen function and $\negl(\cdot)$ is a negligible function.
\item 
We say that an $\rlen$-ensemble ${\cal F}$ is {\sf $(\dlen,
\rlen)$-restricted \corInt} {\em(}or just {\em $\dlen$-\corInt},
for short{\em)}, if for every $\dlen$-respecting
probabilistic polynomial-time machine $M$ 
and every evasive relation $R$ w.r.t. $(\dlen,\rlen)$, 
it holds that
\[\Pr_{s \in \bitset^k}[ x \gets M(s),\;\; (x,f_s(x)) \in R ] = \negl(k)\]
\EI 
{\em Weak $\dlen$-\corInty}\/ is defined analogously by considering
only polynomial-time recognizable $R$'s.
\ED

The rest of this section is dedicated to demonstrating impossibility
results for restricted correlation intractable 
ensembles, in some cases.  We also highlight cases where
existence of restricted correlation intractable               
ensembles is left as an open problem.

\subsection{Negative results for short seeds} 
The proof ideas of \thmref{no-impl1} can be easily applied to rule out 
the existence of certain restricted correlation intractable ensembles
where the seed is too short. 

\BP{~}\label{no-impl2} 
\begin{description}
\item[{(a)}] 
If $\dlen(k) \geq k - O(\log k)$ for infinitely many $k$'s, then there 
exists no ensemble that is $(\dlen,\rlen)$-\corInt, even in the weak sense. 
\item[{(b)}] 
If $\dlen(k)+\rlen(k) \geq k + \omega(\log k)$, there 
exists no ensemble that is $(\dlen,\rlen)$-\corInt. 
\end{description}
\EP
\begin{proof}
The proof of (a) is a straightforward generalization of the proof of 
\thmref{no-impl1}. Actually, we need to consider two cases: the case 
$\dlen(k)\geq k$ and the case $k-O(\log k)\leq\dlen(k)<k$. In the first 
case, we proceed as in the proof of \thmref{no-impl1} 
(except that we define 
$R^{\cal F}\eqdef\{(x,f_s(x)): s \in \bitset^*,x=s0^{\dlen(|s|)-|s|}\}$). 
In the second case, for every ensemble ${\cal F}$, we define the relation 
\[R^{\cal F} \eqdef 
   \left\{(x,f_{xz}(x)) : x,z \in \bitset^*\,,\, |x|=\dlen(|xz|) \right\}
\]
We show that $R^{\cal F}$ is evasive by showing that, 
for every $k\in\N$ and $x\in\bitset^{\dlen(k)}$,  
there exist at most polynomially (in $k$) many $y$'s 
such that $(x,y) \in R^{\cal F}$.
This is the case since $(x,y)\in R^{\cal F}$ implies that there 
exists some $z$ such that $\dlen(|xz|)=|x|$ and $y=f_{xz}(x)$.
But using the case hypothesis we have 
$|x| = \dlen(|xz|) \geq |xz| - O(\log|xz|)$, which implies that 
$|z| = O(\log(|xz|))$ and hence also $|z|=O(\log|x|)$. 
Next, using the other case hypothesis (i.e., $k>\dlen(k)=|x|$), 
we conclude that $|z| = O(\log k)$. 
Therefore, there could be at most polynomially many such $z$'s, 
and so the upper bound on the number of $y$'s paired with $x$ follows.
The evasiveness of $R^{\cal F}$ as well as the assertion that $R^{\cal F}$ 
is polynomial-time computable follow (assuming that the function $\dlen$ 
itself is polynomial-time computable). 
On the other hand, consider the machine $M$ that, on input $s$,
outputs the $\dlen(|s|)$-bit prefix of $s$. 
Then, for every $s\in\bitset^*$, 
we have $(M(s),f_s(M(s)))\in R^{\cal F}$. 

\medskip
For the proof of (b), assume that $\dlen(k) < k$ (for all but finitely 
many $k$'s). We start by defining the ``inverse'' of the $\dlen$ function 
\[ \dlen^{-1}(n) \eqdef {\rm min}\{ k\ :\ \dlen(k) = n \} 
\]
(where, in case there exists no $k$ such that $\dlen(k) = n$, 
we define $\dlen^{-1}(n) = 0$). 
By definition it follows that $k \geq \dlen^{-1}(\dlen(k))$, for all 
$k$'s (because $k$ belongs to the set $\{ k'\ :\ \dlen(k') = \dlen(k)\}$), 
and that $\dlen(\dlen^{-1}(n)) = n$,  whenever there exists some $k$ for 
which $n = \dlen(k)$.  Next we define 
\[{R}^{\cal F}\eqdef\left\{(x,f_{xz}(x))\,:\,x,z\in\bitset^*\,,\,
                                  |x|+|z|=\dlen^{-1}(|x|)\right\}
\]
This relation is well defined since, by the conditions 
on the lengths of $x$ and $z$,
we have $\dlen(|xz|) = \dlen(\dlen^{-1}(|x|)) = |x|$ and so 
the function $f_{xz}$ is indeed defined on the input $x$. 
In case $\dlen(k) \leq k - \omega(\log k)$, this relation may not be 
polynomial-time recognizable. Still, it is evasive w.r.t. $(\dlen,\rlen)$, 
since with security parameter $k$ we have 
for every $x\in\bitset^{\dlen(k)}$ 
\begin{eqnarray*}
  \left|\left\{y\in\bitset^{\rlen(k)}\ : (x,y) \in {R}^{\cal F}\right\}\right| 
  &=&\left| \left\{ f_{xz}(x)\ : |z| = \dlen^{-1}(\dlen(k))-\dlen(k)\right\}
             \ \cap\ \bitset^{\rlen(k)} \right| \\
  &\leq& 2^{\dlen^{-1}(\dlen(k))-\dlen(k)} \\ 
  &\leq& 2^{k-\dlen(k)} 
\end{eqnarray*}
Using $k-\dlen(k)\leq\rlen(k)-\omega(\log k)$,
we conclude that the set of $y$'s paired with $x$ forms a negligible
fraction of $\bitset^{\rlen(k)}$, and so that $R^{\cal F}$ is evasive.
Again, the machine $M$, that on input $s$ outputs 
the $\dlen(|s|)$-bit prefix of $s$, 
satisfies $(M(s),f_s(M(s)))\in R^{\cal F}$, for all $s$'s. 
\end{proof}

\paragraph{Open Problems:}
%-------------------------
Proposition~\ref{no-impl2} still leaves open the question of 
existence of $(\dlen,\rlen)$-restricted \corInt\ ensembles,
for the case $\dlen(k)+\rlen(k)<k+O(\log k)$.\footnote{\label{d-log.fn}
In fact such ensembles do exist in case $k\geq2^{\dlen(k)}\cdot\rlen(k)$
(since the seed may be used to directly specify all the function's values),
but we dismiss this trivial and useless case.}  
We believe that it is interesting to resolve the situation either way: 
Either provide negative results also for the above special case, 
or provide a plausible construction.
Also open is the sub-case where $\dlen(k)+\rlen(k)=k+\omega(\log k)$  
but one considers only {\em weak}\/ $(\dlen,\rlen)$-restricted \corInty. 
(Recall that Case~(b) of Proposition~\ref{no-impl2} 
is proven using relations which are not known 
to be polynomial-time recognizable.) 

We comment that even if restricted \corInt\
ensembles exist, then they are very non-robust constructs.
For example, even if the ensemble ${\cal F} = \{f_s : |s|=k\}_k$ 
is \corInt\ with respect to some length functions $(\dlen,\rlen)$, 
the ensemble that is obtained by applying many independent 
copies of ${\cal F}$ and concatenating the results may not be. 
That is, for $m\!:\!\N\!\to\!\N$, define 
\begin{equation}\label{non-robust.eq1} 
{\cal F}^m \eqdef \{f'_{\langle s_1,...,s_{m(k)}\rangle}
  :|s_1|=\cdots=|s_{m(k)}|=k\}_{k\in\N}\,,
\end{equation}
where, for $\langle x_1,...,x_{m(k)}\rangle\in\bitset^{m(k)\cdot\dlen(k)}$, 
\begin{equation}\label{non-robust.eq2} 
f'_{\langle s_1,...,s_{m(k)}\rangle}(\langle x_1,...,x_{m(k)}\rangle)
  \eqdef \langle f_{s_1}(x_1),....,f_{s_{m(k)}}(x_{m(k)})\rangle\,.
\end{equation}
Then, for sufficiently large $m$ (e.g., $m(k)\geq k/\dlen(k)$ will do),
the ``direct product'' ensemble ${\cal F}^m$ is not \corInt\
(not even in the restricted sense).
That is, 

\BP \label{no-impl3}
Let $\dlen, \rlen: \intfnc$ be length functions
so that $\dlen(k)\leq k$, 
and let $m:\intfnc$ be a polynomially-bounded 
function so that $m(k)\geq k/\dlen(k)$. Let ${\cal F}$ be an 
arbitrary function ensemble, and ${\cal F}^m$ be as defined in 
\eqref{non-robust.eq1} and~(\ref{non-robust.eq2}). Then, ${\cal F}^m$ 
is not \corInt, not even in the $(\dlen^m,\rlen^m)$-restricted 
sense, where $\ell_{\rm in}^m(m(k)\cdot k)\eqdef 
m(k)\cdot\ell_{\rm in}(k)$ and $\ell_{\rm out}^m(m(k)\cdot k)\eqdef 
m(k)\cdot\ell_{\rm out}(k)$.
\EP

\begin{proof}
We assume, for simplicity that $m(k)=k/\dlen(k)$
(and so $\dlen(k)=k/m(k)$ and $\dlen^m(m(k)\cdot k)=k$). 
Given ${\cal F}^m$ as stated, 
we again adapt the proof of \thmref{no-impl1}. 
This time, using $\dlen(k)\leq k$, we define the relation
$$R^{{\cal F}^m} 
   \eqdef \bigcup_k\left\{\,(s,\langle f_{s}(s'),t\rangle)\,:\; 
 |s| = k,\;\; \mbox{\rm $s'$ is the $\dlen(k)$-prefix of $s$},\;\;
 |t| = (m(k)-1)\cdot\rlen(k) \, \right\}$$
Notice that in this definition we have $|s| = \frac{k}{\dlen(k)} 
\cdot \dlen(k) = m(k) \cdot \dlen(k) = \dlen^m(m(k) \cdot k)$, and also 
$|f_s(s')|+|t| = m(k) \cdot \rlen(k) = \rlen^m(m(k) \cdot k)$, so this 
relation is indeed $(\dlen^m,\rlen^m)$-restricted. 

Again, it is easy to see that $R^{\cal F}$ is polynomial-time recognizable, 
and it is evasive since every string $x \in \bitset^k$ is coupled with 
at most a $2^{-\rlen(k)}$ fraction of the possible $(m(k)\cdot\rlen(k))$-bit 
long strings, and $\rlen(k)=\omega(\log k)=\omega(\log(m(k)\cdot k))$. 
(Here we use the hypothesis $m(k)=\poly(k)$.)

On the other hand, consider a (real-life) adversary that given the seed 
${\ov s}=\langle s_1,...,s_{m(k)}\rangle\in\bitset^{m(k)\cdot k}$ 
for the function $f'_{\langle s_1,...,s_{m(k)}\rangle}$, 
sets the input to this function to be equal to $s_1$. 
Denoting the $\dlen(k)$-prefix of $s_1$ 
(equiv., of $\ov s$) by $s_1'$,
it follows that $f_{s_1}(s_1')$ is a prefix 
of $f'_{\langle s_1,...,s_{m(k)}\rangle}(s_1)$
and so $(s_1, f'_{\langle s_1,....,s_{m(k)}\rangle}(s_1)) \in R^{\cal F}$. 
Thus, this real-life adversary violates the (restricted) correlation 
intractability of ${\cal F}^m$. 
\end{proof}

\subsection{Correlation intractability for multiple invocations}
%===========================================
Recall that Proposition~\ref{no-impl2} does not rule out the existence
of  restricted ensembles having seeds that are longer than the sum 
of lengths of their inputs and outputs. However, even for this special 
case the only thing that is not ruled out is a {\em narrow definition} 
that refers to forming rare relationships between a {\em single}\/ 
input-output pair.  In fact, if one generalizes the definition 
of \corInty\ so as to consider evasive relations over unbounded 
sequences of inputs and outputs,  then the negative result in 
\prpref{no-impl2} can be extended for arbitrary $\dlen$ and 
$\rlen$. That is, 

\BD[multi-invocation restricted \corInty]
\label{def-gen-restrict-unpredict}
Let $\dlen,\rlen: \intfnc$ be length functions. 
We consider probabilistic polynomial-time oracle machines 
which on input $1^k$ have oracle access to 
a function $\f:\bitset^{\dlen(k)}\to\bitset^{\rlen(k)}$.
\BI 
\item
A relation $R$ over pairs of binary sequences
is {\sf evasive with respect to $(\dlen,\rlen)$}
{\em(or $(\dlen,\rlen)$-evasive)}
if for any 
probabilistic polynomial-time machine $M$ as above it holds that 
$$
\Pr_\f\left[ (x_1,...,x_m) \gets M^{\f}(1^k)\,;\ \ 
\begin{array}{l}
  |x_1| = \ldots = |x_{m}| = \dlen(k) \\
  \mbox{\rm and } ((x_1,...,x_m),(\f(x_1),...,\f(x_m))\!\in\!R
\end{array}
\;\right]\;\;=\;\;\negl(k)
$$
As usual, $\f:\bitset^{\dlen(k)}\to\bitset^{\rlen(k)}$ is a uniformly
chosen function. 
\item
We say that an $\rlen$-ensemble ${\cal F}$ is 
{\sf $(\dlen,\rlen)$-restricted multi-invocation \corInt} 
{\em(}or just {\em $\dlen$-multi-invocation \corInt}, for short{\em)}, 
if for every $(\dlen,\rlen)$-evasive relation $R$ and 
every probabilistic polynomial-time machine $M$ it holds that
$$
\Pr_{s\in\{0,1\}^k}\left[ (x_1,...,x_m) \gets M(s)\,;\ \ 
\begin{array}{l}
  |x_1| = \ldots = |x_{m}| = \dlen(k) \\
  \mbox{\rm and } ((x_1,...,x_m),(f_s(x_1),...,f_s(x_m))\!\in\!R
\end{array}
\;\right]\;\;=\;\;\negl(k)
$$
\EI 
\ED

\BP \label{no-impl4}
Let $\dlen,\rlen: \intfnc$ be arbitrary length functions, with 
$\dlen(k) \geq 2+\log k$ and $\rlen(k) \geq 1$.  Then there exist no 
$(\dlen,\rlen)$-restricted {\em multi-invocation} {\corInt} function 
ensembles.
\EP 

\begin{proof}
For simplicity, we consider first the case $\rlen(k) \geq 2$. 
Let ${\cal F}$ be an $\rlen$-ensemble.
Adapting the proof of \thmref{no-impl1}, we define the relation
$$
R^{\cal F} \eqdef \bigcup_k \left\{ 
  \left((x_1, \ldots, x_k), (f_s(x_1),\ldots,f_s(x_k))\right) :
\begin{array}{l}
  x_i = (i,s_i), \mbox{ with }s_i \in \{0,1\} \\
  \mbox{\rm and } s=s_1\ldots s_k
\end{array}
\right\}
$$
(Notice that since $\dlen(k)>1+\log k$, the $x_i$'s are indeed in 
the range of the function $f_s$.)  Clearly, this relation 
is polynomial-time recognizable. 
To see that this relation is evasive, notice that for any fixed 
$k$-bit seed $s = s_1 \ldots s_k$, we have 
$$\Pr_{\f}[\f(i,s_i)=f_s(i,s_i)\mbox{ for }i=1\ldots k] 
  = 2^{-\rlen(k)\cdot k}$$ 
Hence, the probability that there exists a seed $s$ 
for which $\f(i,s_i)=f_s(i,s_i)$ holds, for $i=1,...,k$,
is at most $2^k \cdot 2^{-\rlen(k)\cdot k} \leq 2^{-k}$.  
It follows that
$$\Pr_{\f}[\exists x_1,...,x_k\;
           \left((x_1, \ldots, x_k), (\f(x_1),\ldots,\f(x_k))\right)
           \in R^{\cal F}]\leq 2^{-k}$$ 
However, the corresponding multi-invocation restricted \corInty\ condition 
does not hold: 
For any $s = s_1 \ldots s_k\in \bitset^k$, setting $x_i = (i,s_i)$
we get $((x_1,...,x_k),(f_s(x_1),...,f_s(x_k)))\in R^{\cal F}$. 

To rule out the case $\rlen(k) = 1$, we redefine $R^{\cal F}$ so 
that $((x_1,...,x_{2k}),(f_s(x_1),...,f_s(x_{2k})))\in R^{\cal F}$ if 
$x_i=(i,s_i)$ for $i=1,...,k$ and $x_i=(i,0)$ for $i=k+1,...,2k$. 
\end{proof}

\paragraph{Discussion:} 
%%%%%%%%%%%%%%%%%%%%%%%%%%%%%%%%%%%%%%%%%%%%%%%%%%%%%%%%%%%%%%%%%%%
Propositions~\ref{no-impl2}, \ref{no-impl3} and~\ref{no-impl4} 
demonstrate that there is only a very narrow margin in which 
strict correlation-intractability may be used. 
Still, even ensembles that are (strict) correlation-intractable
with respect to relations of a-priori bounded total length 
(of input-output sequences) may be useful in some applications. 
Typically, this may hold in applications 
where number of invocations of the cryptosystem is a-priori bounded
(or where the security of the system depends only on an a-priori 
bounded partial history of invocations; e.g., the current one).
We note that the Fiat-Shamir heuristic 
for transforming interactive identification protocols into
signature schemes~\cite{FiSh86} does not fall into the above category, 
since the function's seed needs to be fixed with the public key, 
and used for signing polynomially many messages, 
where the polynomial is not a-priori known.

\subsection{Correlation intractability for given, polynomial-time
relations}%\label{discuss.sec}
In all our negative results, the evasive relation demonstrating
that a certain function ensemble is not correlation-intractable
is more complex than the function ensemble itself.
A natural restriction on correlation-intractability is to require
that it holds only for relations recognizable within certain
fixed polynomial time bounds (or some fixed space bound), and 
allowing the function ensemble
to have a more complex polynomial-time evaluation algorithm.
We stress that, 
in both the definition of evasiveness and correlation-intractability,
the adversary that generates the inputs to the relation
is allowed arbitrary (polynomial) running time; this time may be larger
than both the time to evaluate the function ensemble 
and the time to evaluate the relation. 
Such a restricted notion of correlation-intractability
may suffice for some applications, and it would be interesting to
determine whether function ensembles satisfying it do exist. 
Partial results in this direction were obtained by Nissim~\cite{kobbi}
and are described next:

\BP[\cite{kobbi}] 
Let $\dlen,\rlen: \intfnc$ be arbitrary length functions, with 
$k\geq\rlen(k)\cdot(\dlen(k)+\omega(\log k))$.% 
\footnote{\
Recall that $(\dlen,\rlen)$-restricted correlation-intractable ensembles
exist for $k\geq2^{\dlen(k)}\cdot\rlen(k)$; see Footnote~\ref{d-log.fn}.}
Then, for every binary relation $R$ that is evasive with respect to 
$(\dlen,\rlen)$ and recognizable in polynomial-time, 
there exists a function ensemble ${\cal F}^R=\{f_s\}$
that is correlation-intractable with respect to $R$;
that is, for every $\dlen$-respecting probabilistic
polynomial-time machine $M$ it holds that
\[\Pr_{s\in\bitset^k}[x\gets M(s),\;\;(x,f_s(x))\in R] = \negl(k)\]
\label{Nissim.prop1} 
\EP 
We note that the postulated construction uses a seed length
that is longer than $\dlen+\rlen$. Thus, this positive result
capitalizes on both restrictions discussed above
(i.e., both the length and the complexity restrictions). 
\medskip 

\begin{proof}
Let $t=\dlen(k)+\omega(\log k)$.
For every seed $s=(s_1,...,s_t)\in\bitset^{t\cdot \rlen(k)}$,
we define $f_s:\bitset^{\dlen(k)}\to\bitset^{\rlen(k)}$
so that $f_{s_1,...,s_t}(x)$ equals $s_i$
if $i$ is the smallest integer such that $(x,s_i)\not\in R$. 
In case $(x,s_i)\in R$ holds for all $i$'s,
we define $f_{s_1,...,s_t}(x)$ arbitrarily.

\def\Gk{S_k}
Let $R(x)\eqdef\{y:(x,y)\in R\}$, 
and $\Gk\eqdef\{x\in\bitset^{\dlen(k)}: |R(x)|\leq 2^{\rlen(k)}/2 \}$ 
(S stands for ``Small image''). 
Since $R$ is evasive, 
it is infeasible to find an $x\in\bitset^{\dlen(k)}$ not in $\Gk$.
Thus, for every probabilistic polynomial-time $M$,
$\Pr_{s\in\bitset^k}[M(s)\not\in \Gk]=\negl(k)$. 
On the other hand, the probability that such $M(s)$
outputs an $x\in \Gk$ so that $(x,f_{s}(x))\in R$
is bounded above by\footnote{\
For the first inequality,
we use the fact that if there exists an $i$ such that $(x,s_i)\not\in R$
then $(x,f_{s}(x))\not\in R$.} 
\begin{eqnarray*} 
\Pr_{s\in\bitset^k}[\exists x\in \Gk\;\;\mbox{\rm s.t. }\;(x,f_{s}(x))\in R]
&\leq&\Pr_{s\in\bitset^k}[\exists x\in \Gk\,\forall i\;\; (x,s_i)\in R] \\
&\leq&|\Gk|\cdot\max_{x\in \Gk}
           \left\{\Pr_{s}[\forall i\;\; (x,s_i)\in R]\right\} \\
&\leq& 2^{\dlen(k)}\cdot(1/2)^t \;=\;  \negl(k)
\end{eqnarray*} 
Combining the two cases, the proposition follows. 
\end{proof}

Considering the notion of multi-invocation correlation-intractability
when restricting the complexity of the relation (and allowing the
function ensemble to be more complex), Nissim has obtained
another impossibility result \cite{kobbi}: 

\BP[\cite{kobbi}]
There exists an evasive relation $R$ that 
is recognizable in polynomial-time
so that no function ensemble ${\cal F}=\{f_s\}$ 
is multi-invocation correlation-intractable with respect to $R$;
that is, for every function ensemble ${\cal F}=\{f_s\}$ 
there exists a polynomial-time machine $M$ such that
$$
\Pr_s\left[(x_1,...,x_t) \gets M(s)\,;\ \;
   ((x_1,...,x_t),(f_s(x_1),...,f_s(x_t))\!\in\!R\;\right]\;\;=\;\;1
$$
Furthermore, for some universal polynomial $p$, 
which is independent of ${\cal F}$, it holds that $t<p(|x_1|)$. 
\label{Nissim.prop2} 
\EP
We stress that the above assertion includes even function ensembles 
that have (polynomial-time) evaluation algorithms of running time 
greater than the time it takes to recognize $t$-tuples of corresponding 
length is the relation. Furthermore, it includes function ensembles
having seeds of length exceeding the total length of pairs in the
relation. 

\medskip\noindent{\bf Proof Sketch:}
We follow the ideas underlying the proof of Theorem~\ref{sign.thm}.
Specifically, using the universal machine $M_{\cal U}$
and the algorithms ($\prv$ and $\ver$) of a CS-proof system,
we consider a relation $R$ that contains pairs of binary sequences, 
so that $((x,\pi,q_1...,q_{m}),(y,\phi,a_1...,a_{m}))\in R$ if 
these strings describe an accepting execution of the CS-verifier 
with respect to machine $M_{\cal U}$. That is, we require that the 
following conditions hold:
\BE
\item 
All the strings $y,\phi,a_1...,a_{m}$ have the same length. Below we 
denote this length by $\rlen$, $|y| =|\phi| =|a_1| =\cdots =|a_{m}| =\rlen$. 

\item 
The string $\pi$ is an alleged CS-proof for the assertion that the 
machine $M_{\cal U}$ accepts the input $(x,y)$ within $t(n)=n^{\log n}$ 
steps, where $n\eqdef|x|+|y|$. 

\item
Given access to an oracle that on queries $q_i$ returns answers $a_i$, 
and given security parameter $n+\rlen$ 
and input $w = (\vect{M_{\cal U}}, (x,y),t(n))$, 
the CS verifier $\ver$ accepts the CS-proof $\pi$ 
after querying the oracle on $q_1 \ldots q_m$ (in this order),
and obtaining the corresponding answers $a_1 \ldots a_m$. 

(Here we use the fact that the verifier is deterministic,
and thus its queries are determined by its input and the answers
to previous queries.) 
\EE

Recall that, by definition, $m$ is bounded by a fixed polynomial
in $n$. In fact, in Micali's construction~\cite{Mi94}, $m$ is 
poly-logarithmic in $n$. 
We comment that, assuming the existence
of suitable collision-intractable hash functions, 
one may obtain $m=1$ (cf.~\cite{NaNi99}.  
In addition, one may need to make some minor modification
in the above construction.)

As in the proof of Theorem~\ref{sign.thm}, 
using the computational soundness of CS-proofs,
it can be shown that the above relation is evasive. 
By the additional efficiency conditions of CS-proofs,
it follows that the relation is recognizable in polynomial-time.
On the other hand, as in the proof of Theorem~\ref{sign.thm}, 
for every function ensemble ${\cal F}^i=\{f^i_s\}$
there exists a polynomial-time adversary $A$,
that on input $s$ produces a sequence $(x,\pi,q_1,...,q_{m})$ so that 
$(
 (x,\pi,q_1,...,q_{m}),(f^i_s(x),f^i_s(\pi),f^i_s(q_1),...,f^i_s(q_{m}))
  ) \in R$. 
This is done as follows:
First $A$ sets $x=\langle i,s\rangle$, $y=f_s^i(x)$, and $n\eqdef|x|+|y|$. 
Next, $A$ constructs a CS-proof that indeed $M_{\cal U}$ accepts $(x,y)$ 
within $n^{\log n}$ steps,
and sets $\pi$ to equal this proof. 
(This step takes time polynomial in the evaluation time of $f_s^i(x)$.)
Note that since $(x,y)$ is indeed accepted by $M_{\cal U}$
(in less than $n^{\log n}$ steps), 
the verifier accept $\pi$ as a proof no matter how the oracle is
determined (since perfect completeness holds). 
Finally, the adversary invokes the verifier
(on input consisting mainly of $(x,y)$ and $\pi$), and (by emulating 
the oracle) determines interactively the oracle queries and answers
of the verifier; that is, for every $j=1,...,m$, the adversary
determines the $j^\th$ query made by the verifier,
sets $q_{j}$ to equal this query, and provides the verifier
with the answer $f^i_s(q_{j})$. \qqed 

%%%%%%%%%%%%%%%%%%%%%%%%%%%%%%%%%%%%%%%%%%%%%%%%%%%%%%%%%%%%%%%%%%%%%
\section{Conclusions}
\label{sec-conclusions}
%%%%%%%%%%%%%%%%%%%%%%%%%%%%%%%%%%%%%%%%%%%%%%%%%%%%%%%%%%%%%%%%%%%%%
The authors  have different opinions regarding the Random Oracle Methodology. 
Rather than trying to strike a mild compromise,
we prefer to present our disagreements in the most controversial form. 

\subsection{Ran's Conclusions}

Real-life cryptographic applications are complex objects. On top of the
``cryptographic core,'' these applications typically involve 
numerous networking protocols,  several other applications,
user-interfaces, and in fact also an entire operating-system.
The security of an application depends on the security of all these
components operating in unison. Thus, in principle, the best way to gain
assurance in the security of a cryptographic application is to analyze
it as a single unit, bones and feathers included.

However, analyzing an entire system is prohibitively complex.
Moreover, we often feel that the ``essence'' of a cryptographic
application
can be presented in a relatively simple way without getting into many 
details, which, we feel, are ``extraneous'' to the actual security.
Consequently, we often make  {\em abstractions} of a cryptographic
application by leaving many details ``outside the model''. 
Nonetheless, some caution is needed when making abstractions:
While good abstractions are important and useful, bad abstractions can 
be dangerous and misleading. Thus, it is crucial to make sure that
one uses a sound abstraction, or one that helps to distinguish between
good and bad applications.

One popular abstraction is to treat computers in a network as interactive 
Turing machines who run one specific (and relatively simple) algorithm, and
assume that delivery of messages is done simply by having one machine
write values on the tapes of another machine. We are then satisfied with
defining and analyzing security of a protocol in this abstract model. In
other words, this abstraction implicitly uses the following methodology 
(which I'll call the ``Interactive Turing machine methodology''): Design 
and analyze a protocol in the ``idealized system'' (i.e., using Turing
machines). Next, come up with an ``implementation'' of the idealized
protocol by adding the components that deal with the networking
protocols, the operating system, the user interfaces, etc. Now, ``hope''
that the implementation is indeed secure.

We widely believe that this methodology is sound, in the sense that
if an idealized protocol is secure then there {\em exist} secure
implementations of it. Furthermore, security of an idealized protocol
is a good predictor for the feasibility of finding a good implementation
to it.  (Of course, finding secure implementations to secure idealized
protocols is a far-from-trivial task, and there is probably no single 
automatic method for securely implementing any idealized protocol. But
this does not undermine the soundness of the 
``Interactive Turing machine methodology''.)

The Random Oracle Methodology is, in essence, another proposed abstraction 
of cryptographic applications. It too proposes to define and analyze 
security of protocols in an idealized model, then perform some 
transformation that is ``outside the formal model'', and now ``hope'' that 
the resulting implementation is secure. At first glance
it looks like a great 
abstraction: It does away with specific implementation issues of
``cryptographic hash functions'' and concentrates on designing protocols
assuming that an ``ideal hash function'' is available. Indeed,
protocols that were 
designed using this methodology are remarkably simple and efficient,
while resisting all known attacks. 

However, as shown in this work, and in sharp contrast to the
``Interactive Turing machine methodology'', the Random Oracle 
Methodology is not sound. Furthermore, it is a bad predictor to the 
security of implementations: Not only do there {\em exist}
idealized protocols that have no secure implementations, practically 
{\em any} idealized protocol can be slightly ``tweaked'' so that the
tweaked protocol remains just  as secure in the idealized model, 
but has no secure implementations. This leaves us no choice but
concluding that, in spite of its apparent successes, the Random Oracle 
Methodology  is a bad abstraction of protocols for the purpose of analyzing
security.

\paragraph{The loss of reductions to hard problems.}
The above argument should provide sufficient motivation to be wary of
security analyses in the \ROM.
Nonetheless, let us highlight the following additional 
disturbing aspect of such analysis.

One of the great contributions of complexity-based modern cryptography, 
developed in the past quarter of a century, is the ability to base the
security of many varied protocols on a small number of well-defined
and well-studied complexity assumptions. Furthermore, typically the
proof of security of a protocol provides us with  a method for
transforming adversary that breaks the security of the said protocol
into an adversary that refutes one of the well-studied assumptions. 
In light of our inability to prove security of protocols from scratch,
this methodology provides us with the ``next best'' evidence for the
security of protocols.

The Random Oracle Methodology does away with these advantages.
Assume that an idealized protocol $A$ is proven secure in the
\ROM\ based on, say, the Diffie-Hellman assumption, and that someone
comes up with a way to break  % some implementation of $A$, or even 
{\em any implementation} of $A$. This does
not necessarily mean that it is now possible to break Diffie-Hellman!
Consequently, the reducibility of the security of $A$ to the hardness of
Diffie-Hellman is void. This brings us back to a situation
where the security of each protocol is a ``stand-alone'' problem
and is, in essence, unrelated to the hardness of known problems.

\paragraph{Possible alternative directions.}
In spite of its shortcomings, the Random Oracle Methodology seems 
to generate simple and efficient protocols against which no attacks are 
known. One possible direction towards providing formal 
justification for some
of these protocols is to identify useful, 
special-purpose properties of the random oracle, which can be also 
provided by a fully specified function (or function ensemble)  
and so yield secure implementations of certain useful ideal systems. 
First steps in this direction were taken in \cite{Ca97,CMR98,GHR99}.
Hopefully, future works will push this direction further.

\subsection{Oded's Conclusions}
My starting point is that within the domain of science, 
every deduction requires a rigorous justification.%
\footnote{\ 
This does not disallow creative steps committed
in the course of research, without proper justification. 
Such unjustified steps are the fuel of progress. 
What I refer to are claims that are supposed to reflect valid facts.
Such claims should be fully justified, or offered as conjectures.} 
In contrast, unjustified deductions should not be allowed;
especially not in a subtle research area such as Cryptography. 
Furthermore, one should refrain from making statements that are
likely to mislead the listener/reader, such as claiming a
result in a restricted model while creating the impression 
that it holds also in a  less restricted model. 
The presentation of such a result should clearly state 
the restrictions under which it holds, 
and refrain from creating the impression that the result
extends also to a case where these restrictions are waived
(unless this is indeed true (and one can prove it)). 
Needless to say, it is perfectly {\sc ok} to conjecture that 
a restricted result extends also to a case when these restrictions 
are waived, but the stature of such a statement (as a conjecture) 
should be clear.

The above abstract discussion directly applies to security in the \ROM.
Deducing that the security of a scheme in the \ROM\/ 
means anything about the security of its implementations,
without proper justification, is clearly wrong.%
\footnote{\ Using the \ROM\/ as a justification to the feasibility
of meeting some security requirements is even ``more wrong''.}  
This should have been clear also before the current work.
It should have also been clear that no proper justification 
of deduction from security in the \ROM\/ to security
of implementations has been given (so far).
The contributions of the current work are two-fold:
\BE
\item %Correlation Intractability {sec-unpredict}
This work uncovers inherent difficulties in the project of 
providing conditions that would allow 
(justifiable) deduction from security in
the \ROM\/ to security of implementations. 
Such a project could have proceeded by identifying properties 
that characterize proofs of security in the \ROM,
and (justifiably) deducing that the such schemes maintain their security 
when implemented with ensembles satisfying these properties. 
The problem with this project is that correlation intractability
should have been (at the very least) one of these properties,
but (as we show) no function ensemble can satisfy it. 
\item %Failure {sec-insecure}
As stated above, deducing that the security of a scheme in the \ROM\/
means anything about the security of its implementations,
without proper justification, is clearly wrong.
The current work presents concrete examples 
in which this unjustified deduction leads to wrong conclusions.
That is, it is shown that not only that unjustified deduction
regarding the \ROM\/ {\em {\sc may} lead to wrong conclusions},
but rather than in some cases {\sc indeed} this unjustified
deduction {\em {\sc does} lead to wrong conclusions}.
Put in other words, if one needs a concrete demonstration 
of the dangers of unjustified deduction when applied to the \ROM,
then this work provides it. % I personally did not need it. 
\EE

\paragraph{The bottom-line:}
It should be clear that the Random Oracle Methodology
is not sound; that is, the mere fact that a scheme is secure in
the \ROM\/ cannot be taken as evidence (or indication) to
the security of (possible) implementations of this scheme. 
Does this mean that the \ROM\/ is useless?
Not necessarily: it may be useful as a test-bed (or as a sanity check).% 
\footnote{\
This explains the fact the Random Oracle
Methodology is in fact used in practice. 
In also explains why many reasonable schemes, 
the security of which is still an open problem,
are secure in the \ROM: 
good suggestions should be expected to pass a sanity check.} 
Indeed, if the scheme does not perform well on the test-bed
(resp., fails the sanity check) then it should be dumped. 
But one should {\em not}\/ draw wrong conclusions from the mere fact
that a scheme performs well on the test-bed
(resp., passes the sanity check). 
In summary, the Random Oracle Methodology is actually 
a method for ruling out {\em some}\/ insecure designs, 
but this method is not ``complete'' 
(i.e., it may fail to rule out insecure designs).%
\footnote{\ 
Would I, personally, endorse this method 
is a different question. My answer is very much time-sensitive:
Given the current misconceptions regarding the \ROM, 
I would suggest not to include, in currently published work, 
proofs of security in the \ROM.
My rationale is that the dangers of misconceptions
(regarding such proofs) seem to out-weight 
the gain of demonstrating that the scheme passed a sanity check.
I hope that in the future such misconceptions will be less
prevailing, at which time it would be indeed recommended to
report on the result of a sanity check.} 

\subsection{Shai's Conclusions}
The negative results in this work (and in particular Theorems~\ref{sign.thm} 
and~\ref{encrypt.thm}) leave me with an uneasy feeling: adopting 
the view that a good theory should be able to explain ``the real 
world'', I would have liked theoretical results that explain the 
apparent success of the random oracle methodology in devising useful, 
seemingly secure, cryptographic schemes. (Indeed, this was one of the 
original motivations for this work.) 
Instead, in this work we show that security of cryptographic schemes 
in the {\ROM} does not necessarily imply security in ``the real world''. 
Trying to resolve this apparent mismatch, one may come up with several 
different explanations. Some of those are discussed below:
\BI
\item 
The current success of this methodology is due to pure luck: all the 
current schemes that are proven secure in the {\ROM}, happen to be secure 
also in the ``real world'' for no reason.   However, our ``common sense'' 
and sense of esthetics must lead us to reject such explanation. 

\item 
The current apparent success is a mirage: some of the schemes that are 
proven secure in the {\ROM} are not really secure, and attacks on them 
may be discovered in the future. 

This explanation seems a little more attractive than the previous one. 
After all, a security proof in the {\ROM} eliminates a broad class of 
potential attacks (i.e., the ones that would work also in the {\ROM}), 
and in many cases it seems that attacks of this type are usually the 
ones that are easier to find.  Hence, it makes sense that if there exists 
a ``real life'' attack on a scheme which is secure in the {\ROM}, it may 
be harder -- and take longer -- to find this attack. 
Still, the more time passes without published attacks against ``real 
life'' schemes which are proven secure in the {\ROM}, the less likely 
this explanation would become. 

\item 
Another possible explanation is that the random oracle methodology 
works for the current published schemes, due to some specific features 
of these schemes that we are yet to identify.  That is, maybe it is 
possible to identify interesting classes of schemes, for which security 
in the {\ROM} implies the existence of a secure implementation.%
\footnote{One particularly silly example are schemes that do not use the 
oracle.  Another, more interesting example, are schemes that only use 
the ``perfect one-way'' property of the oracle; see \cite{Ca97,CMR98}. 
}\  

Identifying such interesting classes, and proving the above implication, 
is an important -- and seemingly hard -- research direction.  (In fact, 
it even seems to be hard to identify classes of schemes for which this 
implication makes a reasonable computational assumption.) 
To appreciate the difficulty in proceeding towards this goal, recall 
that the techniques in the work can be used to ``tweak'' almost any 
cryptographic scheme into one which is secure in the {\ROM} but has 
no secure implementation.  Hence, any classification as above must 
be refined enough to separate the original scheme (for which we want 
to prove that security in the {\ROM} implies security in the real world) 
from the ``tweaked'' one (for which this implication does not hold). 
\EI

My bottom line is that at the present time, the random oracle methodology 
seems to be a very useful ``engineering tool'' for devising schemes. 
As a practical matter, I would much rather see today's standards built 
around schemes which are proven secure in the {\ROM}, than around 
schemes for which no such proofs exist. 
The results in this paper, however, must serve as a warning that 
security proof in the {\ROM} can never be thought of as the end of 
the road.  Proofs in the {\ROM} are useful in that they eliminate a 
broad class of attacks, but they do not imply that other attacks cannot 
be found. 

In terms of scientific research, our works clearly demonstrate that 
the random oracle methodology is not sound in general.  My feeling is 
that our understanding of the relations between the {\ROM} and the 
``real world'' is very limited.  As I said above, it would be very 
interesting to identify special cases in which the random oracle 
methodology is sound.  Similarly, it would be interesting to see other, 
``less artificial'', examples where this methodology fails.

%%%%%%%%%%%%%%%%%%%%%%%%%%%%%%%%%%%%%%%%%%%%%%%%%%%%%%%%%%%%%%%%%%%
\section*{Acknowledgments}
We wish to thank Silvio Micali for enlightening discussions. 
We thank Clemens Holenstein for uncovering a flaw in the 
proof of \prpref{no-impl4} in an earlier version of this paper. 
Special thanks to Kobbi Nissim for telling us about his related results
(i.e., Propositions~\ref{Nissim.prop1} and~\ref{Nissim.prop2}) 
and permitting us to include them in this write-up. 
%%%%%%%%%%%%%%%%%%%%%%%%%%%%%%%%%%%%%%%%%%%%%%%%%%%%%%%%%%%%%%%%%%%

\bibliographystyle{abbrv}

\begin{thebibliography}{10}

\bibitem{BeRo93}
M.~Bellare and P.~Rogaway.
\newblock Random oracles are practical: a paradigm for designing efficient
  protocols.
\newblock In {\em 1st Conference on Computer and Communications Security},
  pages 62--73. ACM, 1993.

\bibitem{BeRo96}
M.~Bellare and P.~Rogaway.
\newblock The exact security of digital signatures: How to sign with rsa and
  rabin.
\newblock In {\em Advances in Cryptology - {EUROCRYPT}'96}, volume 1070 of {\em
  Lecture Notes in Computer Science}, pages 399--416. Springer-Verlag, 1996.

\bibitem{BlMi84}
M.~Blum and S.~Micali.
\newblock How to generate cryptographically strong sequences of pseudo-random
  bits.
\newblock {\em {SIAM Journal on Computing}}, 13:850--864, 1984.

\bibitem{Ca97}
R.~Canetti.
\newblock Towards realizing random oracles: Hash functions that hide all
  partial information.
\newblock In {\em Advances in Cryptology - {CRYPTO}'97}, volume 1294 of {\em
  Lecture Notes in Computer Science}, pages 455--469. Springer-Verlag, 1997.

\bibitem{CGH98}
R. Canetti, O. Goldreich and S. Halevi. 
The Random Oracle Methodology, Revisited.
\newblock In {\em Proceedings of the 30th Annual ACM Symposium on the Theory of
  Computing},  Dallas, TX, May 1998. ACM.

\bibitem{CMR98}
R.~Canetti, D.~Micciancio and O.~Reingold.
\newblock Perfectly one-way probabilistic hashing.
\newblock In {\em Proceedings of the 30th Annual ACM Symposium on the Theory of
  Computing}, pages 131--140, Dallas, TX, May 1998. ACM.

\bibitem{Da87}
I.~Damg{\aa}rd.
\newblock Collision free hash functions and public key signature schemes.
\newblock In {\em Advances in Cryptology - {EUROCRYPT}'87}, volume 304 of {\em
  Lecture Notes in Computer Science}, pages 203--216. Springer-Verlag, 1987.

\bibitem{magic}
C.~Dwork, M.~Naor, O.~Reingold, and L.~Stockmeyer.
\newblock Magic functions.
\newblock In {\em 40th Annual Symposium on Foundations of Computer Science},
  pages 523--534. IEEE, 1999.

\bibitem{FiSh86}
A.~Fiat and A.~Shamir.
\newblock How to prove yourself. practical solutions to identification and
  signature problems.
\newblock In {\em Advances in Cryptology - {CRYPTO}'86}, volume 263 of {\em
  Lecture Notes in Computer Science}, pages 186--189. Springer-Verlag, 1986.

\bibitem{GHR99}
R. Gennaro, S. Halevi and T. Rabin. Secure Hash-and-Sign Signatures Without
the Random Oracle.
\newblock In {\em Advances in Cryptology - {EUROCRYPT}'99}, volume 1592 of {\em
  Lecture Notes in Computer Science}, pages 123--139. Springer-Verlag, 1999.

\bibitem{G93} O.~Goldreich.
\newblock A Uniform Complexity Treatment of Encryption and Zero-Knowledge.
\newblock {\em Journal of Cryptology}, Vol.~6, No.~1, pages 21--53, 1993.

\bibitem{G99} O.~Goldreich.
\newblock {\em Encryption Schemes -- fragments of a chapter}.
\newblock December 1999.
\newblock Available from
          {\tt http://www.wisdom.weizmann.ac.il/$\sim$oded/foc-book.html}

\bibitem{ggm86}
O.~Goldreich, S.~Goldwasser, and S.~Micali.
\newblock How to construct random functions.
\newblock {\em Journal of the ACM}, 33(4):210--217, 1986.

\bibitem{GK}
O.~Goldreich, and H.~Krawczyk.
\newblock On the Composition of Zero-Knowledge Proof Systems.
\newblock {\em SIAM Journal on Computing}, 25(1):169--192, 1996. 

\bibitem{GoMi84}
S.~Goldwasser and S.~Micali.
\newblock Probabilistic encryption.
\newblock {\em Journal of Computer and System Sciences}, 28(2):270--299, April
  1984.

\bibitem{gmr88}
S.~Goldwasser, S.~Micali, and R.~Rivest.
\newblock A digital signature scheme secure against adaptive chosen-message
  attacks.
\newblock {\em SIAM Journal of Computing}, 17(2):281--308, Apr. 1988.

\bibitem{GuQu88}
L.~Guillou and J.~Quisquater.
\newblock A practical zero-knowledge protocol fitted to security
  microprocessors minimizing both transmission and memory.
\newblock In {\em Advances in Cryptology - {EUROCRYPT}'88}, volume 330 of {\em
  Lecture Notes in Computer Science}, pages 123--128. Springer-Verlag, 1988.

\bibitem{HaTa99}
S.~Hada and T.~Tanaka.
\newblock A relationship between one-wayness and correlation intractability.
\newblock In {\em The 2nd
  International Workshop on Practice and Theory in Public Key Cryptography
  (PKC'99)}, volume 1560 of {\em Lecture Notes in Computer Science}, pages
  82--96, Japan, March 1999. Springer-Verlag.

\bibitem{ImRu89}
R.~Impagliazzo and S.~Rudich.
\newblock Limits on the provable consequences of one-way permutations.
\newblock In {\em Proceedings of the 21st Annual {ACM} Symposium on Theory of
  Computing}, pages 44--61, Seattle, WA, May 1989. ACM.

\bibitem{Ki92}
J.~Kilian.
\newblock A note on efficient zero-knowledge proofs and arguments.
\newblock In {\em Proceedings of the 24th Annual ACM Symposium on the Theory of
  Computing}, pages 723--732. ACM, May 1992.

\bibitem{Mi94}
S.~Micali.
\newblock {CS} proofs.
\newblock In {\em 35th Annual Symposium on Foundations of Computer Science
  ("FOCS'94")}, pages 436--453. IEEE, 1994.

\bibitem{NaNi99}
M.~Naor and K.~Nissim.
\newblock Computationally sound proofs: Reducing the number of random oracle
  calls.
\newblock manuscript, 1999.

\bibitem{ny89}
M.~Naor and M.~Yung.
\newblock Universal one-way hash functions and their cryptographic
  applications.
\newblock In {\em Proceedings of the 21st Annual {ACM} Symposium on Theory of
  Computing}, pages 33--43, 1989.

\bibitem{kobbi} K.~Nissim. 
\newblock Two results regarding correlation intractability. 
\newblock Manuscript, 1999.

\bibitem{Ok92}
T.~Okamoto.
\newblock Provably secure and practical identification scheme and corresponding
  signature scheme.
\newblock In {\em Advances in Cryptology - {CRYPTO}'92}, volume 740 of {\em
  Lecture Notes in Computer Science}, pages 31--53. Springer-Verlag, 1992.

\bibitem{PoSt96}
D.~Pointcheval and J.~Stern.
\newblock Security proofs for signature schemes.
\newblock In {\em Advances in Cryptology - {EUROCRYPT}'96}, volume 1070 of {\em
  Lecture Notes in Computer Science}, pages 387--398. Springer-Verlag, 1996.

\bibitem{ro90}
J.~Rompel.
\newblock One-way functions are necessary and sufficient for secure signatures.
\newblock In {\em Proceedings of the 22nd Annual {ACM} Symposium on Theory of
  Computing}, pages 387--394, 1990.

\bibitem{Sc91}
C.~Schnorr.
\newblock Efficient signature generation by smart cards.
\newblock {\em {Journal of Cryptology}}, 4(3):161--174, 1991.

\bibitem{yao82a}
A.~C. Yao.
\newblock Theory and applications of trapdoor functions.
\newblock In {\em 23rd Annual Symposium on Foundations of Computer Science},
  pages 80--91. IEEE, Nov. 1982.

\end{thebibliography}

\end{document}